\documentstyle[aps,pra]{revtex} 
\newcommand{\e}[1]{\text{e}^{#1}}

\newcommand{\hide}[1]{}
\newcommand{\showlabel}[1]{\eqnum{#1}\label{#1}}
\newcommand{\ext}{_{\text{ext}}}

\newcommand{\dg}{^{\dag}}

\input{prepictex}
\input{pictex}
\input{postpictex}
\newcommand{\solid}{%
\setplotsymbol ({\scriptsize .}  <0ex,0.02ex> )
\setsolid
}
\newcommand{\dotted}{%
\setplotsymbol ({\tiny .}  <0ex,0.02ex> )
\setdots <0.3333333ex>
}
\newcommand{\linePT}[5]{%
\put(0,0){\beginpicture
\setcoordinatesystem units <1ex,1ex>
\setlinear 
\plot #1 #2 #3 #4 /
\endpicture}
\subp{#3}{#4}{#5}
}

\newcommand{\arrPT}[2]{%
\put(0,0){\beginpicture
\setcoordinatesystem units <1ex,1ex>
\arrow <1ex> [0.4, 0.8] from #1 to #2 
\endpicture}
}

\newcommand{\arcPT}[3]{%
\put(0,0){\beginpicture
\setcoordinatesystem units <1ex,1ex>
\circulararc #1 degrees from #2 center at #3
\endpicture}
}

\newcommand{\subp}[3]{%
\put(#1,#2){\begin{picture}(0,0)#3\end{picture}}%
}

\newcommand{\vrt}[2]{%
\thicklines\put(#1,#2){\circle*{1}}}
\newcommand{\lnd}[6]{%
\put(#1,#2){\begin{picture}(0,0)
\thicklines \put(0,0){\line(1,0){#3}}
\put(0.5,1){\footnotesize #4}
\put(#3,0){\begin{picture}(0,0)#6%
\put(-2.5,1){\footnotesize #5}
\end{picture}}\end{picture}}}
\newcommand{\tarr}[3]{%
\thinlines 
\put(#1,#2){\vector(1,0){#3}}
}
\newcommand{\vframe}[5]{%
\left \{
\begin{picture}(0.8,#2)(0,0)\end{picture}
\begin{picture}(#1,#2)(0,-0.55)
\put(#3,#4){\begin{picture}(0,0)(0,0)#5%
\end{picture}}\end{picture}
\begin{picture}(0.8,#2)(0,0)\end{picture}
\right \}%
}
\newcommand{\dframe}[3]{%
\left \{
\hspace{1ex}
\begin{picture}(#1,#2)(0,-0.55)#3\end{picture}
\hspace{1ex}
\right \}%
}

\newcommand{\lndrs}[6]{%
\put(#1,#2){\begin{picture}(0,0)#6%
\thicklines \put(0,0){\line(1,0){#3}}
\put(0,0){\circle*{1}}\put(#3,0){\circle*{1}}
\put(0.5,-2.5){\footnotesize #4}
\put(#3,0){\begin{picture}(0,0)
\put(-2.5,-2.5){\footnotesize #5}
\end{picture}}\end{picture}}}

\newcommand{\txt}[3]{\put(#1,#2){\footnotesize #3}}
\newcommand{\showPT}[2]{}
\newcommand{\figPTa}{%
\dframe{4.5}{3.25}{\subp{0}{0.25}{%
\arrPT{2.4 0}{2.5 0}
\linePT{0}{0}{4.}{0}{}
\subp{4.}{0}{%
\txt{-0.5}{0.5}{\makebox[0ex][l]{$t$}}
}
\tarr{0.25}{-2.}{4.}
}
}
}
\showPT{\figPTa}{figPTa}
\newcommand{\figPTb}{%
\dframe{4.5}{3.25}{\subp{0}{0.25}{%
\arrPT{1.6 0}{1.5 0}
\linePT{0}{0}{4.}{0}{}
\subp{4.}{0}{%
\txt{-0.5}{0.5}{\makebox[0ex][l]{$t$}}
}
\tarr{0.25}{-2.}{4.}
}
}
}
\showPT{\figPTb}{figPTb}
\newcommand{\figPTc}{%
\dframe{9.5}{3.5}{\subp{1.}{0.}{%
\txt{0.5}{0.5}{\makebox[0ex][r]{$t'$}}
\arrPT{4.4 0}{4.5 0}
\linePT{0}{0}{8.}{0}{}
\subp{8.}{0}{%
\txt{-0.5}{0.5}{\makebox[0ex][l]{$t$}}
}
\tarr{1.75}{-2.}{4.}
}
}
}
\showPT{\figPTc}{figPTc}
\newcommand{\figPTd}{%
\dframe{9.5}{3.5}{\subp{1.}{0.}{%
\txt{0.5}{0.5}{\makebox[0ex][r]{$t'$}}
\arrPT{3.6 0}{3.5 0}
\linePT{0}{0}{8.}{0}{}
\subp{8.}{0}{%
\txt{-0.5}{0.5}{\makebox[0ex][l]{$t$}}
}
\tarr{1.75}{-2.}{4.}
}
}
}
\showPT{\figPTd}{figPTd}
\newcommand{\figPTe}{%
\dframe{10.5}{3.}{\subp{1.5}{0.5}{%
\txt{-0.5}{0}{\makebox[0ex][r]{$x'$}}
\vrt{4.}{0}
\dotted
\arrPT{2.4 0}{2.5 0}
\linePT{0}{0}{4.}{0}{}
\arrPT{6.4 0}{6.5 0}
\linePT{4.}{0}{8.}{0}{}
\subp{8.}{0}{%
\txt{0.5}{0}{\makebox[0ex][l]{$x$}}
}
\tarr{1.75}{-2.5}{4.}
}
}
}
\showPT{\figPTe}{figPTe}
\newcommand{\figPTf}{%
\dframe{10.5}{3.}{\subp{1.5}{0.5}{%
\txt{-0.5}{0}{\makebox[0ex][r]{$x'$}}
\vrt{4.}{0}
\dotted
\arrPT{1.6 0}{1.5 0}
\linePT{0}{0}{4.}{0}{}
\arrPT{5.6 0}{5.5 0}
\linePT{4.}{0}{8.}{0}{}
\subp{8.}{0}{%
\txt{0.5}{0}{\makebox[0ex][l]{$x$}}
}
\tarr{1.75}{-2.5}{4.}
}
}
}
\showPT{\figPTf}{figPTf}
\newcommand{\figPTg}{%
\dframe{9.}{2.75}{\subp{0}{0.75}{%
\vrt{4.}{0}
\solid
\arrPT{2.4 0}{2.5 0}
\linePT{0}{0}{4.}{0}{}
\dotted
\arrPT{6.4 0}{6.5 0}
\linePT{4.}{0}{8.}{0}{}
\subp{8.}{0}{%
\txt{0.5}{0}{\makebox[0ex][l]{$x$}}
}
\tarr{2.5}{-2.5}{4.}
}
}
}
\showPT{\figPTg}{figPTg}
\newcommand{\figPTh}{%
\dframe{9.}{2.75}{\subp{0}{0.75}{%
\vrt{4.}{0}
\solid
\arrPT{1.6 0}{1.5 0}
\linePT{0}{0}{4.}{0}{}
\dotted
\arrPT{5.6 0}{5.5 0}
\linePT{4.}{0}{8.}{0}{}
\subp{8.}{0}{%
\txt{0.5}{0}{\makebox[0ex][l]{$x$}}
}
\tarr{2.5}{-2.5}{4.}
}
}
}
\showPT{\figPTh}{figPTh}
\newcommand{\figPTi}{%
\dframe{20.}{3.}{\subp{11.}{0.5}{%
\solid
\arrPT{-9.4 0}{-9.5 0}
\linePT{-11.}{0}{-7.}{0}{}
\subp{-7.}{0}{%
\txt{0.5}{0}{\makebox[0ex][l]{$x''$}}
}
\txt{-0.5}{0}{\makebox[0ex][r]{$x'$}}
\vrt{4.}{0}
\dotted
\arrPT{2.4 0}{2.5 0}
\linePT{0}{0}{4.}{0}{}
\arrPT{6.4 0}{6.5 0}
\linePT{4.}{0}{8.}{0}{}
\subp{8.}{0}{%
\txt{0.5}{0}{\makebox[0ex][l]{$x$}}
}
\tarr{-3.}{-2.5}{4.}
}
}
}
\showPT{\figPTi}{figPTi}
\newcommand{\figPTj}{%
\dframe{13.}{2.75}{\subp{0}{0.75}{%
\vrt{4.}{0}
\solid
\arrPT{2.4 0}{2.5 0}
\linePT{0}{0}{4.}{0}{}
\arrPT{8.4 0}{8.5 0}
\linePT{4.}{0}{12.}{0}{}
\subp{12.}{0}{%
\txt{0.5}{0}{\makebox[0ex][l]{$x$}}
}
\tarr{4.5}{-2.5}{4.}
}
}
}
\showPT{\figPTj}{figPTj}
\newcommand{\figPTk}{%
\dframe{13.}{2.75}{\subp{0}{0.75}{%
\vrt{4.}{0}
\solid
\arrPT{1.6 0}{1.5 0}
\linePT{0}{0}{4.}{0}{}
\arrPT{7.6 0}{7.5 0}
\linePT{4.}{0}{12.}{0}{}
\subp{12.}{0}{%
\txt{0.5}{0}{\makebox[0ex][l]{$x$}}
}
\tarr{4.5}{-2.5}{4.}
}
}
}
\showPT{\figPTk}{figPTk}
\newcommand{\figPTl}{%
\dframe{17.}{2.75}{\subp{0}{0.75}{%
\vrt{4.}{0}
\solid
\arrPT{2.4 0}{2.5 0}
\linePT{0}{0}{4.}{0}{}
\arrPT{8.4 0}{8.5 0}
\linePT{4.}{0}{12.}{0}{}
\subp{12.}{0}{%
\vrt{0}{0}
\dotted
\arrPT{2.4 0}{2.5 0}
\linePT{0}{0}{4.}{0}{}
\subp{4.}{0}{%
\txt{0.5}{0}{\makebox[0ex][l]{$x$}}
}
}
\tarr{6.5}{-2.5}{4.}
}
}
}
\showPT{\figPTl}{figPTl}
\newcommand{\figPTm}{%
\dframe{21.}{2.75}{\subp{0}{0.75}{%
\vrt{4.}{0}
\solid
\arrPT{2.4 0}{2.5 0}
\linePT{0}{0}{4.}{0}{}
\arrPT{8.4 0}{8.5 0}
\linePT{4.}{0}{12.}{0}{}
\subp{12.}{0}{%
\vrt{0}{0}
\arrPT{4.4 0}{4.5 0}
\linePT{0}{0}{8.}{0}{}
\subp{8.}{0}{%
\txt{0.5}{0}{\makebox[0ex][l]{$x$}}
}
}
\tarr{8.5}{-2.5}{4.}
}
}
}
\showPT{\figPTm}{figPTm}
\newcommand{\figPTn}{%
\dframe{29.}{2.75}{\subp{0}{0.75}{%
\vrt{4.}{0}
\solid
\arrPT{2.4 0}{2.5 0}
\linePT{0}{0}{4.}{0}{}
\arrPT{8.4 0}{8.5 0}
\linePT{4.}{0}{12.}{0}{}
\subp{12.}{0}{%
\vrt{0}{0}
\arrPT{4.4 0}{4.5 0}
\linePT{0}{0}{8.}{0}{}
\subp{8.}{0}{%
\vrt{0}{0}
\arrPT{4.4 0}{4.5 0}
\linePT{0}{0}{8.}{0}{}
\subp{8.}{0}{%
\txt{0.5}{0}{\makebox[0ex][l]{$x$}}
}
}
}
\tarr{12.5}{-2.5}{4.}
}
}
}
\showPT{\figPTn}{figPTn}
\newcommand{\figPTo}{%
\dframe{4.5}{2.75}{\subp{0}{0.25}{%
\solid\setdashpattern <1.ex, 0.5ex>
\linePT{0}{0}{4.}{0}{}
\subp{4.}{0}{%
\txt{-0.5}{0.5}{\makebox[0ex][l]{$x$}}
}
\tarr{0.25}{-2.}{4.}
}
}
}
\showPT{\figPTo}{figPTo}
\newcommand{\figPTp}{%
\dframe{4.5}{2.75}{\subp{0.5}{0.25}{%
\solid\setdashpattern <1.ex, 0.5ex>
\txt{-0.5}{0.5}{\makebox[0ex][l]{$x$}}
\linePT{0}{0}{4.}{0}{}
\tarr{-0.25}{-2.}{4.}
}
}
}
\showPT{\figPTp}{figPTp}
\newcommand{\figPTq}{%
\dframe{4.5}{2.5}{\subp{0}{1.}{%
\setdashpattern <1ex, 0.5ex>
\linePT{0}{0}{4.}{0}{}
\subp{4.}{0}{%
\vrt{0}{0}
}
\tarr{0.25}{-2.5}{4.}
}
}
}
\showPT{\figPTq}{figPTq}
\newcommand{\figPTr}{%
\dframe{4.5}{2.5}{\subp{0.5}{1.}{%
\setdashpattern <1ex, 0.5ex>
\vrt{0}{0}
\linePT{0}{0}{4.}{0}{}
\tarr{-0.25}{-2.5}{4.}
}
}
}
\showPT{\figPTr}{figPTr}
\newcommand{\figPTs}{%
\dframe{5.5}{4.5}{\subp{0.5}{-1.5}{%
\setdashpattern <1ex, 0.5ex>
\linePT{0}{3.}{4.}{3.}{}
\subp{4.}{3.}{%
\txt{-0.5}{0.5}{\makebox[0ex][l]{$x'$}}
}
\txt{-0.5}{0.5}{\makebox[0ex][l]{$x$}}
\linePT{0}{0}{4.}{0}{}
\tarr{0.25}{-2.}{4.}
}
}
}
\showPT{\figPTs}{figPTs}
\newcommand{\figPTt}{%
\dframe{9.5}{3.}{\subp{0.5}{0.}{%
\solid
\txt{-0.5}{0.5}{\makebox[0ex][l]{$x$}}
\linePT{0}{0}{8.}{0}{}
\subp{8.}{0}{%
\txt{-0.5}{0.5}{\makebox[0ex][l]{$x'$}}
}
\tarr{2.25}{-2.}{4.}
}
}
}
\showPT{\figPTt}{figPTt}
\newcommand{\figPTu}{%
\dframe{6.}{6.}{\subp{0.5}{-3.}{%
\solid\setdashpattern <1.ex, 0.5ex>
\linePT{0}{6.}{4.}{6.}{}
\subp{4.}{6.}{%
\txt{-0.5}{0.5}{\makebox[0ex][l]{$x''$}}
}
\linePT{0}{3.}{4.}{3.}{}
\subp{4.}{3.}{%
\txt{-0.5}{0.5}{\makebox[0ex][l]{$x'$}}
}
\txt{-0.5}{0.5}{\makebox[0ex][l]{$x$}}
\linePT{0}{0}{4.}{0}{}
\tarr{0.5}{-2.}{4.}
}
}
}
\showPT{\figPTu}{figPTu}
\newcommand{\figPTv}{%
\dframe{10.}{4.5}{\subp{0.5}{-1.5}{%
\solid\setdashpattern <1.ex, 0.5ex>
\linePT{4.}{3.}{8.}{3.}{}
\subp{8.}{3.}{%
\txt{-0.5}{0.5}{\makebox[0ex][l]{$x''$}}
}
\solid
\txt{-0.5}{0.5}{\makebox[0ex][l]{$x$}}
\linePT{0}{0}{8.}{0}{}
\subp{8.}{0}{%
\txt{-0.5}{0.5}{\makebox[0ex][l]{$x'$}}
}
\tarr{2.5}{-2.}{4.}
}
}
}
\showPT{\figPTv}{figPTv}
\newcommand{\figPTw}{%
\dframe{10.}{4.5}{\subp{0.5}{-1.5}{%
\solid\setdashpattern <1.ex, 0.5ex>
\linePT{4.}{3.}{8.}{3.}{}
\subp{8.}{3.}{%
\txt{-0.5}{0.5}{\makebox[0ex][l]{$x'$}}
}
\solid
\txt{-0.5}{0.5}{\makebox[0ex][l]{$x$}}
\linePT{0}{0}{8.}{0}{}
\subp{8.}{0}{%
\txt{-0.5}{0.5}{\makebox[0ex][l]{$x''$}}
}
\tarr{2.5}{-2.}{4.}
}
}
}
\showPT{\figPTw}{figPTw}
\newcommand{\figPTx}{%
\dframe{4.5}{2.75}{\subp{0}{0.25}{%
\solid
\linePT{0}{0}{4.}{0}{}
\subp{4.}{0}{%
\txt{-0.5}{0.5}{\makebox[0ex][l]{$x$}}
}
\tarr{0.25}{-2.}{4.}
}
}
}
\showPT{\figPTx}{figPTx}
\newcommand{\figPTy}{%
\dframe{5.}{4.5}{\subp{0}{-1.5}{%
\solid\setdashpattern <1.ex, 0.5ex>
\linePT{0}{0}{4.}{0}{}
\subp{4.}{0}{%
\txt{-0.5}{0.5}{\makebox[0ex][l]{$x$}}
}
\linePT{0}{3.}{4.}{3.}{}
\subp{4.}{3.}{%
\txt{-0.5}{0.5}{\makebox[0ex][l]{$x'$}}
}
\tarr{0.5}{-2.}{4.}
}
}
}
\showPT{\figPTy}{figPTy}
\newcommand{\figPTz}{%
\dframe{5.}{4.5}{\subp{0}{-1.5}{%
\solid\setdashpattern <1.ex, 0.5ex>
\linePT{0}{0}{4.}{0}{}
\subp{4.}{0}{%
\txt{-0.5}{0.5}{\makebox[0ex][l]{$x$}}
}
\solid
\linePT{0}{3.}{4.}{3.}{}
\subp{4.}{3.}{%
\txt{-0.5}{0.5}{\makebox[0ex][l]{$x'$}}
}
\tarr{0.5}{-2.}{4.}
}
}
}
\showPT{\figPTz}{figPTz}
\newcommand{\figPTaa}{%
\dframe{5.}{4.25}{\subp{0}{-1.25}{%
\solid\setdashpattern <1.ex, 0.5ex>
\linePT{0}{0}{4.}{0}{}
\subp{4.}{0}{%
\txt{-0.5}{0.5}{\makebox[0ex][l]{$x'$}}
}
\solid
\linePT{0}{3.}{4.}{3.}{}
\subp{4.}{3.}{%
\txt{-0.5}{0.5}{\makebox[0ex][l]{$x$}}
}
\tarr{0.5}{-2.}{4.}
}
}
}
\showPT{\figPTaa}{figPTaa}
\newcommand{\figPTab}{%
\dframe{5.}{4.5}{\subp{0}{-1.5}{%
\solid
\linePT{0}{0}{4.}{0}{}
\subp{4.}{0}{%
\txt{-0.5}{0.5}{\makebox[0ex][l]{$x$}}
}
\linePT{0}{3.}{4.}{3.}{}
\subp{4.}{3.}{%
\txt{-0.5}{0.5}{\makebox[0ex][l]{$x'$}}
}
\tarr{0.5}{-2.}{4.}
}
}
}
\showPT{\figPTab}{figPTab}
\newcommand{\figPTac}{%
\dframe{4.5}{2.5}{\subp{0}{1.}{%
\solid
\linePT{0}{0}{4.}{0}{}
\subp{4.}{0}{%
\vrt{0}{0}
}
\tarr{0.25}{-2.5}{4.}
}
}
}
\showPT{\figPTac}{figPTac}
\newcommand{\figPTad}{%
\dframe{4.5}{4.}{\subp{0}{-0.5}{%
\solid\setdashpattern <1.ex, 0.5ex>
\linePT{0}{0}{4.}{0}{}
\subp{4.}{0}{%
\vrt{0}{0}
}
\linePT{0}{3.}{4.}{3.}{}
\subp{4.}{3.}{%
\vrt{0}{0}
}
\tarr{0.25}{-2.5}{4.}
}
}
}
\showPT{\figPTad}{figPTad}
\newcommand{\figPTae}{%
\dframe{4.5}{4.}{\subp{0}{-0.5}{%
\solid\setdashpattern <1.ex, 0.5ex>
\linePT{0}{0}{4.}{0}{}
\subp{4.}{0}{%
\vrt{0}{0}
}
\solid
\linePT{0}{3.}{4.}{3.}{}
\subp{4.}{3.}{%
\vrt{0}{0}
}
\tarr{0.25}{-2.5}{4.}
}
}
}
\showPT{\figPTae}{figPTae}
\newcommand{\figPTaf}{%
\dframe{4.5}{4.}{\subp{0}{-0.5}{%
\solid
\linePT{0}{0}{4.}{0}{}
\subp{4.}{0}{%
\vrt{0}{0}
}
\linePT{0}{3.}{4.}{3.}{}
\subp{4.}{3.}{%
\vrt{0}{0}
}
\tarr{0.25}{-2.5}{4.}
}
}
}
\showPT{\figPTaf}{figPTaf}
\newcommand{\figPTag}{%
\dframe{11.}{2.5}{\subp{0.5}{1.}{%
\solid\setdashpattern <1.ex, 0.5ex>
\vrt{0}{0}
\linePT{0}{0}{4.}{0}{}
\linePT{6.}{0}{10.}{0}{}
\subp{10.}{0}{%
\vrt{0}{0}
}
\tarr{3.}{-2.5}{4.}
}
}
}
\showPT{\figPTag}{figPTag}
\newcommand{\figPTah}{%
\dframe{5.}{4.}{\subp{0.5}{-0.5}{%
\solid\setdashpattern <1.ex, 0.5ex>
\vrt{0}{0}
\linePT{0}{0}{4.}{0}{}
\solid
\linePT{0}{3.}{4.}{3.}{}
\subp{4.}{3.}{%
\vrt{0}{0}
}
\tarr{0.}{-2.5}{4.}
}
}
}
\showPT{\figPTah}{figPTah}
\newcommand{\figPTai}{%
\dframe{9.}{2.5}{\subp{0.5}{1.}{%
\solid
\vrt{0}{0}
\linePT{0}{0}{8.}{0}{}
\subp{8.}{0}{%
\vrt{0}{0}
}
\tarr{2.}{-2.5}{4.}
}
}
}
\showPT{\figPTai}{figPTai}
\newcommand{\figPTaj}{%
\dframe{9.}{3.25}{\subp{0.5}{0.25}{%
\dotted
\txt{-0.5}{0.5}{\makebox[0ex][l]{$x'$}}
\vrt{4.}{0}
\linePT{0}{0}{8.}{0}{}
\subp{8.}{0}{%
\txt{-0.5}{0.5}{\makebox[0ex][l]{$x$}}
}
\tarr{2.}{-2.5}{4.}
}
}
}
\showPT{\figPTaj}{figPTaj}
\newcommand{\figPTak}{%
\dframe{8.}{2.5}{\subp{0}{1.}{%
\solid\setdashpattern <1.ex, 0.5ex>
\vrt{4.}{0}
\linePT{0}{0}{8.}{0}{}
\tarr{2.}{-2.5}{4.}
}
}
}
\showPT{\figPTak}{figPTak}
\newcommand{\figPTal}{%
\dframe{12.5}{2.5}{\subp{0}{1.}{%
\solid
\vrt{4.}{0}
\vrt{12.}{0}
\linePT{0}{0}{12.}{0}{}
\tarr{4.25}{-2.5}{4.}
}
}
}
\showPT{\figPTal}{figPTal}
\newcommand{\figPTam}{%
\dframe{17.}{2.5}{\subp{0.5}{1.}{%
\solid
\vrt{0}{0}
\vrt{8.}{0}
\vrt{16.}{0}
\linePT{0}{0}{16.}{0}{}
\tarr{6.}{-2.5}{4.}
}
}
}
\showPT{\figPTam}{figPTam}
\newcommand{\figPTan}{%
\dframe{20.5}{2.5}{\subp{0}{1.}{%
\solid
\vrt{4.}{0}
\vrt{12.}{0}
\vrt{20.}{0}
\linePT{0}{0}{20.}{0}{}
\tarr{8.25}{-2.5}{4.}
}
}
}
\showPT{\figPTan}{figPTan}
\newcommand{\figPTao}{%
\dframe{25.}{2.5}{\subp{0.5}{1.}{%
\solid
\vrt{0}{0}
\vrt{8.}{0}
\vrt{16.}{0}
\vrt{24.}{0}
\linePT{0}{0}{24.}{0}{}
\tarr{10.}{-2.5}{4.}
}
}
}
\showPT{\figPTao}{figPTao}
\newcommand{\figPTap}{%
\dframe{4.5}{2.75}{\subp{0}{0.25}{%
\solid
\linePT{0}{0}{4.}{0}{}
\subp{4.}{0}{%
\txt{-0.5}{0.5}{\makebox[0ex][l]{$x$}}
}
\tarr{0.25}{-2.}{4.}
}
}
}
\showPT{\figPTap}{figPTap}
\newcommand{\figPTaq}{%
\dframe{12.5}{3.}{\subp{0}{0.5}{%
\solid
\vrt{4.}{0}
\linePT{0}{0}{12.}{0}{}
\subp{12.}{0}{%
\txt{-0.5}{0.5}{\makebox[0ex][l]{$x$}}
}
\tarr{4.25}{-2.5}{4.}
}
}
}
\showPT{\figPTaq}{figPTaq}
\newcommand{\figPTar}{%
\dframe{20.5}{3.}{\subp{0}{0.5}{%
\solid
\vrt{4.}{0}
\vrt{12.}{0}
\linePT{0}{0}{20.}{0}{}
\subp{20.}{0}{%
\txt{-0.5}{0.5}{\makebox[0ex][l]{$x$}}
}
\tarr{8.25}{-2.5}{4.}
}
}
}
\showPT{\figPTar}{figPTar}
\newcommand{\figPTas}{%
\dframe{9.}{3.}{\subp{0.5}{0.5}{%
\solid
\vrt{0}{0}
\linePT{0}{0}{8.}{0}{}
\subp{8.}{0}{%
\txt{-0.5}{0.5}{\makebox[0ex][l]{$x$}}
}
\tarr{2.}{-2.5}{4.}
}
}
}
\showPT{\figPTas}{figPTas}
\newcommand{\figPTat}{%
\dframe{17.}{3.}{\subp{0.5}{0.5}{%
\solid
\vrt{0}{0}
\vrt{8.}{0}
\linePT{0}{0}{16.}{0}{}
\subp{16.}{0}{%
\txt{-0.5}{0.5}{\makebox[0ex][l]{$x$}}
}
\tarr{6.}{-2.5}{4.}
}
}
}
\showPT{\figPTat}{figPTat}
\newcommand{\figPTau}{%
\dframe{25.}{3.}{\subp{0.5}{0.5}{%
\solid
\vrt{0}{0}
\vrt{8.}{0}
\vrt{16.}{0}
\linePT{0}{0}{24.}{0}{}
\subp{24.}{0}{%
\txt{-0.5}{0.5}{\makebox[0ex][l]{$x$}}
}
\tarr{10.}{-2.5}{4.}
}
}
}
\showPT{\figPTau}{figPTau}
\newcommand{\figPTav}{%
\dframe{11.9641}{4.}{\subp{11.4641}{1.}{%
\solid
\vrt{0}{0}
\linePT{0}{0}{-8.}{0}{}
\subp{-8.}{0}{%
\vrt{0}{0}
\linePT{0}{0}{-3.4641}{2.}{}
\linePT{0}{0}{-3.4641}{-2.}{}
}
\tarr{-7.48205}{-4.}{4.}
}
}
}
\showPT{\figPTav}{figPTav}
\newcommand{\figPTaw}{%
\dframe{18.8923}{8.5}{\subp{18.3923}{1.}{%
\solid
\vrt{0}{0}
\linePT{0}{0}{-8.}{0}{}
\subp{-8.}{0}{%
\vrt{0}{0}
\linePT{0}{0}{-6.9282}{4.}{}
\subp{-6.9282}{4.}{%
\vrt{0}{0}
\linePT{0}{0}{-3.4641}{2.}{}
\linePT{0}{0}{-3.4641}{-2.}{}
}
\linePT{0}{0}{-6.9282}{-4.}{}
\subp{-6.9282}{-4.}{%
\vrt{0}{0}
\linePT{0}{0}{-3.4641}{2.}{}
\linePT{0}{0}{-3.4641}{-2.}{}
}
}
\tarr{-10.9462}{-8.}{4.}
}
}
}
\showPT{\figPTaw}{figPTaw}
\newcommand{\figPTax}{%
\dframe{17.}{3.25}{\subp{16.5}{0.25}{%
\solid
\vrt{0}{0}
\linePT{0}{0}{-8.}{0}{}
\subp{-8.}{0}{%
\vrt{0}{0}
\linePT{0}{0}{-3.4641}{2.}{}
\linePT{0}{0}{-8.}{0}{}
\subp{-8.}{0}{%
\vrt{0}{0}
}
}
\tarr{-10.}{-2.5}{4.}
}
}
}
\showPT{\figPTax}{figPTax}
\newcommand{\figPTay}{%
\dframe{18.8923}{7.25}{\subp{18.3923}{1.75}{%
\solid
\vrt{0}{0}
\linePT{0}{0}{-8.}{0}{}
\subp{-8.}{0}{%
\vrt{0}{0}
\linePT{0}{0}{-6.9282}{4.}{}
\subp{-6.9282}{4.}{%
\vrt{0}{0}
}
\linePT{0}{0}{-6.9282}{-4.}{}
\subp{-6.9282}{-4.}{%
\vrt{0}{0}
\linePT{0}{0}{-3.4641}{2.}{}
\linePT{0}{0}{-3.4641}{-2.}{}
}
}
\tarr{-10.9462}{-8.}{4.}
}
}
}
\showPT{\figPTay}{figPTay}
\newcommand{\figPTaz}{%
\dframe{15.9282}{6.5}{\subp{15.4282}{1.}{%
\solid
\vrt{0}{0}
\linePT{0}{0}{-8.}{0}{}
\subp{-8.}{0}{%
\vrt{0}{0}
\linePT{0}{0}{-6.9282}{4.}{}
\subp{-6.9282}{4.}{%
\vrt{0}{0}
}
\linePT{0}{0}{-6.9282}{-4.}{}
\subp{-6.9282}{-4.}{%
\vrt{0}{0}
}
}
\tarr{-9.4641}{-6.5}{4.}
}
}
}
\showPT{\figPTaz}{figPTaz}

\newcommand{\figPTba}{%
\dframe{9.}{3.}{\subp{0.5}{0.}{%
\solid
\txt{-0.5}{0.5}{\makebox[0ex][l]{$x'$}}
\linePT{0}{0}{8.}{0}{}
\subp{8.}{0}{%
\txt{-0.5}{0.5}{\makebox[0ex][l]{$x$}}
}
\tarr{2.}{-2.}{4.}
}
}
}
\showPT{\figPTba}{figPTba}
\newcommand{\figPTbb}{%
\dframe{17.}{3.25}{\subp{0.5}{0.25}{%
\solid
\txt{-0.5}{0.5}{\makebox[0ex][l]{$x'$}}
\vrt{8.}{0}
\linePT{0}{0}{16.}{0}{}
\subp{16.}{0}{%
\txt{-0.5}{0.5}{\makebox[0ex][l]{$x$}}
}
\tarr{6.}{-2.5}{4.}
}
}
}
\showPT{\figPTbb}{figPTbb}
\newcommand{\figPTbc}{%
\dframe{25.}{3.25}{\subp{0.5}{0.25}{%
\solid
\txt{-0.5}{0.5}{\makebox[0ex][l]{$x'$}}
\vrt{8.}{0}
\vrt{16.}{0}
\linePT{0}{0}{24.}{0}{}
\subp{24.}{0}{%
\txt{-0.5}{0.5}{\makebox[0ex][l]{$x$}}
}
\tarr{10.}{-2.5}{4.}
}
}
}
\showPT{\figPTbc}{figPTbc}
\newcommand{\figPTbd}{%
\dframe{8.5}{3.}{\subp{0}{0.5}{%
\solid
\linePT{0}{0}{4.}{0}{}
\dotted
\vrt{4.}{0}
\linePT{4.}{0}{8.}{0}{}
\subp{8.}{0}{%
\txt{-0.5}{0.5}{\makebox[0ex][l]{$x$}}
}
\tarr{2.25}{-2.5}{4.}
}
}
}
\showPT{\figPTbd}{figPTbd}
\newcommand{\figPTbe}{%
\dframe{5.}{3.}{\subp{0.5}{0.5}{%
\dotted
\vrt{0}{0}
\linePT{0}{0}{4.}{0}{}
\subp{4.}{0}{%
\txt{-0.5}{0.5}{\makebox[0ex][l]{$x$}}
}
\tarr{0.}{-2.5}{4.}
}
}
}
\showPT{\figPTbe}{figPTbe}
\newcommand{\figPTbf}{%
\dframe{5.9641}{4.5}{\subp{0.5}{1.}{%
\dotted
\vrt{0}{0}
\linePT{0}{0}{3.4641}{2.}{}
\subp{3.4641}{2.}{%
\txt{0.5}{-0.5}{\makebox[0ex][l]{$x$}}
}
\linePT{0}{0}{3.4641}{-2.}{}
\subp{3.4641}{-2.}{%
\txt{0.5}{-0.5}{\makebox[0ex][l]{$x'$}}
}
\tarr{0.482051}{-4.5}{4.}
}
}
}
\showPT{\figPTbf}{figPTbf}
\newcommand{\figPTbg}{%
\dframe{4.}{4.}{\subp{0.517949}{1.}{%
\solid\setdashpattern <1.ex, 0.5ex>
\vrt{0}{0}
\linePT{0}{0}{3.4641}{2.}{}
\linePT{0}{0}{3.4641}{-2.}{}
\tarr{-0.517949}{-4.}{4.}
}
}
}
\showPT{\figPTbg}{figPTbg}
\newcommand{\figPTbh}{%
\dframe{7.9282}{6.5}{\subp{0.5}{1.}{%
\solid
\vrt{0}{0}
\linePT{0}{0}{6.9282}{4.}{}
\subp{6.9282}{4.}{%
\vrt{0}{0}
}
\linePT{0}{0}{6.9282}{-4.}{}
\subp{6.9282}{-4.}{%
\vrt{0}{0}
}
\tarr{1.4641}{-6.5}{4.}
}
}
}
\showPT{\figPTbh}{figPTbh}
\newcommand{\figPTbi}{%
\dframe{9.4282}{6.5}{\subp{0.5}{1.}{%
\solid
\vrt{0}{0}
\linePT{0}{0}{6.9282}{4.}{}
\subp{6.9282}{4.}{%
\txt{0.5}{-0.5}{\makebox[0ex][l]{$x$}}
}
\linePT{0}{0}{6.9282}{-4.}{}
\subp{6.9282}{-4.}{%
\txt{0.5}{-0.5}{\makebox[0ex][l]{$x'$}}
}
\tarr{2.2141}{-6.5}{4.}
}
}
}
\showPT{\figPTbi}{figPTbi}
\newcommand{\figPTbj}{%
\dframe{9.9641}{4.5}{\subp{4.5}{1.}{%
\dotted
\vrt{0}{0}
\linePT{0}{0}{-4.}{0}{}
\subp{-4.}{0}{%
\txt{-0.5}{0.5}{\makebox[0ex][l]{$x''$}}
}
\linePT{0}{0}{3.4641}{2.}{}
\subp{3.4641}{2.}{%
\txt{0.5}{-0.5}{\makebox[0ex][l]{$x$}}
}
\linePT{0}{0}{3.4641}{-2.}{}
\subp{3.4641}{-2.}{%
\txt{0.5}{-0.5}{\makebox[0ex][l]{$x'$}}
}
\tarr{-1.51795}{-4.5}{4.}
}
}
}
\showPT{\figPTbj}{figPTbj}
\newcommand{\figPTbk}{%
\dframe{7.4641}{4.}{\subp{4.}{1.}{%
\solid\setdashpattern <1.ex, 0.5ex>
\vrt{0}{0}
\linePT{0}{0}{-4.}{0}{}
\linePT{0}{0}{3.4641}{2.}{}
\linePT{0}{0}{3.4641}{-2.}{}
\tarr{-2.26795}{-4.}{4.}
}
}
}
\showPT{\figPTbk}{figPTbk}
\newcommand{\figPTbl}{%
\dframe{11.4282}{6.5}{\subp{4.}{1.}{%
\solid
\vrt{0}{0}
\linePT{0}{0}{-4.}{0}{}
\linePT{0}{0}{6.9282}{4.}{}
\subp{6.9282}{4.}{%
\vrt{0}{0}
}
\linePT{0}{0}{6.9282}{-4.}{}
\subp{6.9282}{-4.}{%
\vrt{0}{0}
}
\tarr{-0.285898}{-6.5}{4.}
}
}
}
\showPT{\figPTbl}{figPTbl}
\newcommand{\figPTbm}{%
\dframe{14.8564}{8.5}{\subp{0.5}{-1.}{%
\solid
\vrt{0}{0}
\linePT{0}{0}{6.9282}{4.}{}
\subp{6.9282}{4.}{%
\vrt{0}{0}
\linePT{0}{0}{6.9282}{4.}{}
\subp{6.9282}{4.}{%
\vrt{0}{0}
}
\linePT{0}{0}{6.9282}{-4.}{}
\subp{6.9282}{-4.}{%
\vrt{0}{0}
}
}
\linePT{0}{0}{6.9282}{-4.}{}
\subp{6.9282}{-4.}{%
\vrt{0}{0}
}
\tarr{4.9282}{-6.5}{4.}
}
}
}
\showPT{\figPTbm}{figPTbm}
\newcommand{\figPTbn}{%
\dframe{18.3564}{8.5}{\subp{4.}{-1.}{%
\solid
\vrt{0}{0}
\linePT{0}{0}{-4.}{0}{}
\linePT{0}{0}{6.9282}{4.}{}
\subp{6.9282}{4.}{%
\vrt{0}{0}
\linePT{0}{0}{6.9282}{4.}{}
\subp{6.9282}{4.}{%
\vrt{0}{0}
}
\linePT{0}{0}{6.9282}{-4.}{}
\subp{6.9282}{-4.}{%
\vrt{0}{0}
}
}
\linePT{0}{0}{6.9282}{-4.}{}
\subp{6.9282}{-4.}{%
\vrt{0}{0}
}
\tarr{3.1782}{-6.5}{4.}
}
}
}
\showPT{\figPTbn}{figPTbn}
\newcommand{\figPTbo}{%
\dframe{15.4457}{9.23616}{\subp{0.5}{1.}{%
\solid
\vrt{0}{0}
\linePT{0}{0}{6.9282}{4.}{}
\subp{6.9282}{4.}{%
\vrt{0}{0}
\linePT{0}{0}{7.51754}{2.73616}{}
\subp{7.51754}{2.73616}{%
\vrt{0}{0}
}
\linePT{0}{0}{7.51754}{-2.73616}{}
\subp{7.51754}{-2.73616}{%
\vrt{0}{0}
}
}
\linePT{0}{0}{6.9282}{-4.}{}
\subp{6.9282}{-4.}{%
\vrt{0}{0}
\linePT{0}{0}{7.51754}{2.73616}{}
\subp{7.51754}{2.73616}{%
\vrt{0}{0}
}
\linePT{0}{0}{7.51754}{-2.73616}{}
\subp{7.51754}{-2.73616}{%
\vrt{0}{0}
}
}
\tarr{5.22287}{-9.23616}{4.}
}
}
}
\showPT{\figPTbo}{figPTbo}
\newcommand{\figPTbp}{%
\dframe{17.}{4.3094}{\subp{0.5}{1.}{%
\solid
\vrt{0}{0}
\arcPT{-120.}{0 0}{4. -2.3094}
\subp{8.}{0}{}
\arcPT{120.}{0 0}{4. 2.3094}
\subp{8.}{0}{%
\vrt{0}{0}
\linePT{0}{0}{8.}{0}{}
\subp{8.}{0}{%
\vrt{0}{0}
}
}
\tarr{6.}{-4.3094}{4.}
}
}
}
\showPT{\figPTbp}{figPTbp}
\newcommand{\figPTbr}{%
\dframe{20.5}{4.3094}{\subp{4.}{1.}{%
\solid
\vrt{0}{0}
\linePT{0}{0}{-4.}{0}{}
\arcPT{-120.}{0 0}{4. -2.3094}
\subp{8.}{0}{}
\arcPT{120.}{0 0}{4. 2.3094}
\subp{8.}{0}{%
\vrt{0}{0}
\linePT{0}{0}{8.}{0}{}
\subp{8.}{0}{%
\vrt{0}{0}
}
}
\tarr{4.25}{-4.3094}{4.}
}
}
}
\showPT{\figPTbr}{figPTbr}
\newcommand{\figPTbs}{%
\dframe{41.}{6.6188}{\subp{0.5}{1.}{%
\solid
\vrt{0}{0}
\linePT{0}{0}{40.}{0}{}
\subp{40.}{0}{%
\vrt{0}{0}
}
\vrt{8.}{0}
\arcPT{-120.}{8. 0}{16. -4.6188}
\subp{24.}{0}{%
\vrt{0}{0}
}
\vrt{16.}{0}
\arcPT{120.}{16. 0}{24. 4.6188}
\subp{32.}{0}{%
\vrt{0}{0}
}
\tarr{18.}{-6.6188}{4.}
}
}
}
\showPT{\figPTbs}{figPTbs}
\newcommand{\figPTbt}{%
\dframe{23.9282}{6.5}{\subp{0.5}{1.}{%
\solid
\vrt{0}{0}
\linePT{0}{0}{6.9282}{4.}{}
\subp{6.9282}{4.}{%
\vrt{0}{0}
\linePT{0}{0}{16.}{0}{}
\subp{16.}{0}{%
\vrt{0}{0}
}
\linePT{0}{0}{8.}{-8.}{}
\subp{8.}{-8.}{%
\vrt{0}{0}
}
}
\linePT{0}{0}{6.9282}{-4.}{}
\subp{6.9282}{-4.}{%
\vrt{0}{0}
\linePT{0}{0}{16.}{0}{}
\subp{16.}{0}{%
\vrt{0}{0}
}
\linePT{0}{0}{8.}{8.}{}
\subp{8.}{8.}{%
\vrt{0}{0}
}
}
\tarr{9.4641}{-6.5}{4.}
}
}
}
\showPT{\figPTbt}{figPTbt}
\newcommand{\figPTbu}{%
\dframe{22.8564}{6.5}{\subp{0.5}{1.}{%
\solid
\vrt{0}{0}
\linePT{0}{0}{6.9282}{4.}{}
\subp{6.9282}{4.}{%
\vrt{0}{0}
\linePT{0}{0}{8.}{0}{}
\subp{8.}{0}{%
\vrt{0}{0}
}
\linePT{0}{0}{6.9282}{-4.}{}
\subp{6.9282}{-4.}{%
\vrt{0}{0}
}
}
\linePT{0}{0}{6.9282}{-4.}{}
\subp{6.9282}{-4.}{%
\vrt{0}{0}
\linePT{0}{0}{8.}{0}{}
\subp{8.}{0}{%
\vrt{0}{0}
}
\linePT{0}{0}{6.9282}{4.}{}
\subp{6.9282}{4.}{%
\vrt{0}{0}
\linePT{0}{0}{8.}{0}{}
\subp{8.}{0}{%
\vrt{0}{0}
}
}
}
\tarr{8.9282}{-6.5}{4.}
}
}
}
\showPT{\figPTbu}{figPTbu}
\newcommand{\figPTbv}{%
\dframe{10.}{4.5}{\subp{0.5}{-1.5}{%
\solid
\linePT{4.}{3.}{8.}{3.}{}
\subp{8.}{3.}{%
\txt{-0.5}{0.5}{\makebox[0ex][l]{$x''$}}
}
\txt{-0.5}{0.5}{\makebox[0ex][l]{$x$}}
\linePT{0}{0}{8.}{0}{}
\subp{8.}{0}{%
\txt{-0.5}{0.5}{\makebox[0ex][l]{$x'$}}
}
\tarr{2.5}{-2.}{4.}
}
}
}
\showPT{\figPTbv}{figPTbv}
\newcommand{\figPTbw}{%
\dframe{10.}{4.5}{\subp{0.5}{-1.5}{%
\solid
\linePT{4.}{3.}{8.}{3.}{}
\subp{8.}{3.}{%
\txt{-0.5}{0.5}{\makebox[0ex][l]{$x'$}}
}
\txt{-0.5}{0.5}{\makebox[0ex][l]{$x$}}
\linePT{0}{0}{8.}{0}{}
\subp{8.}{0}{%
\txt{-0.5}{0.5}{\makebox[0ex][l]{$x''$}}
}
\tarr{2.5}{-2.}{4.}
}
}
}
\showPT{\figPTbw}{figPTbw}
\newcommand{\figPTbx}{%
\dframe{6.}{7.5}{\subp{0.5}{-1.5}{%
\solid\setdashpattern <1.ex, 0.5ex>
\linePT{0}{6.}{4.}{6.}{}
\subp{4.}{6.}{%
\txt{-0.5}{0.5}{\makebox[0ex][l]{$x''$}}
}
\linePT{0}{3.}{4.}{3.}{}
\subp{4.}{3.}{%
\txt{-0.5}{0.5}{\makebox[0ex][l]{$x'''$}}
}
\txt{-0.5}{0.5}{\makebox[0ex][l]{$x$}}
\linePT{0}{0}{4.}{0}{}
\txt{-0.5}{-2.5}{\makebox[0ex][l]{$x'$}}
\linePT{0}{-3.}{4.}{-3.}{}
\tarr{0.5}{-5.}{4.}
}
}
}
\showPT{\figPTbx}{figPTbx}
\newcommand{\figPTby}{%
\dframe{10.}{4.5}{\subp{0.5}{-1.5}{%
\solid
\txt{-0.5}{3.5}{\makebox[0ex][l]{$x$}}
\linePT{0}{3.}{8.}{3.}{}
\subp{8.}{3.}{%
\txt{-0.5}{0.5}{\makebox[0ex][l]{$x''$}}
}
\txt{-0.5}{0.5}{\makebox[0ex][l]{$x'$}}
\linePT{0}{0}{8.}{0}{}
\subp{8.}{0}{%
\txt{-0.5}{0.5}{\makebox[0ex][l]{$x'''$}}
}
\tarr{2.5}{-2.}{4.}
}
}
}
\showPT{\figPTby}{figPTby}
\newcommand{\figPTbz}{%
\dframe{10.}{4.5}{\subp{0.5}{-1.5}{%
\solid
\txt{-0.5}{3.5}{\makebox[0ex][l]{$x$}}
\linePT{0}{3.}{8.}{3.}{}
\subp{8.}{3.}{%
\txt{-0.5}{0.5}{\makebox[0ex][l]{$x'''$}}
}
\txt{-0.5}{0.5}{\makebox[0ex][l]{$x'$}}
\linePT{0}{0}{8.}{0}{}
\subp{8.}{0}{%
\txt{-0.5}{0.5}{\makebox[0ex][l]{$x''$}}
}
\tarr{2.5}{-2.}{4.}
}
}
}
\showPT{\figPTbz}{figPTbz}
\newcommand{\figPTca}{%
\dframe{11.}{4.}{\subp{0.5}{-0.5}{%
\solid\setdashpattern <1.ex, 0.5ex>
\vrt{0}{0}
\linePT{0}{0}{4.}{0}{}
\linePT{6.}{0}{10.}{0}{}
\subp{10.}{0}{%
\vrt{0}{0}
}
\vrt{0}{3.}
\linePT{0}{3.}{4.}{3.}{}
\linePT{6.}{3.}{10.}{3.}{}
\subp{10.}{3.}{%
\vrt{0}{0}
}
\tarr{3.}{-2.5}{4.}
}
}
}
\showPT{\figPTca}{figPTca}
\newcommand{\figPTcb}{%
\dframe{9.}{4.}{\subp{0.5}{-0.5}{%
\solid
\vrt{0}{0}
\linePT{0}{0}{8.}{0}{}
\subp{8.}{0}{%
\vrt{0}{0}
}
\vrt{0}{3.}
\linePT{0}{3.}{8.}{3.}{}
\subp{8.}{3.}{%
\vrt{0}{0}
}
\tarr{2.}{-2.5}{4.}
}
}
}
\showPT{\figPTcb}{figPTcb}
\newcommand{\figPTcc}{%
\dframe{9.}{4.}{\subp{0.5}{-0.5}{%
\solid
\vrt{0}{0}
\linePT{0}{0}{8.}{3.}{}
\subp{8.}{3.}{%
\vrt{0}{0}
}
\vrt{0}{3.}
\linePT{0}{3.}{8.}{0}{}
\subp{8.}{0}{%
\vrt{0}{0}
}
\tarr{2.}{-2.5}{4.}
}
}
}
\showPT{\figPTcc}{figPTcc}
\newcommand{\figPTcd}{%
\dframe{5.}{3.5}{\subp{0.5}{0.5}{%
\vrt{0}{0}
\dotted
\arrPT{2.4 0}{2.5 0}
\linePT{0}{0}{4.}{0}{}
\subp{4.}{0}{%
\txt{-0.5}{0.5}{\makebox[0ex][l]{$t$}}
}
\tarr{0.}{-2.5}{4.}
}
}
}
\showPT{\figPTcd}{figPTcd}
\newcommand{\figPTce}{%
\dframe{5.}{3.5}{\subp{0.5}{0.5}{%
\vrt{0}{0}
\dotted
\arrPT{1.6 0}{1.5 0}
\linePT{0}{0}{4.}{0}{}
\subp{4.}{0}{%
\txt{-0.5}{0.5}{\makebox[0ex][l]{$t$}}
}
\tarr{0.}{-2.5}{4.}
}
}
}
\showPT{\figPTce}{figPTce}
\newcommand{\figPTcf}{%
\dframe{5.}{3.5}{\subp{0.5}{0.5}{%
\dotted
\txt{-0.5}{0.5}{\makebox[0ex][l]{$t$}}
\arrPT{2.4 0}{2.5 0}
\linePT{0}{0}{4.}{0}{}
\subp{4.}{0}{%
\vrt{0}{0}
}
\tarr{0.}{-2.5}{4.}
}
}
}
\showPT{\figPTcf}{figPTcf}
\newcommand{\figPTcg}{%
\dframe{5.}{3.5}{\subp{0.5}{0.5}{%
\dotted
\txt{-0.5}{0.5}{\makebox[0ex][l]{$t$}}
\arrPT{1.6 0}{1.5 0}
\linePT{0}{0}{4.}{0}{}
\subp{4.}{0}{%
\vrt{0}{0}
}
\tarr{0.}{-2.5}{4.}
}
}
}
\showPT{\figPTcg}{figPTcg}
\newcommand{\figPTch}{%
\dframe{10.3284}{6.07843}{\subp{5.82843}{0.25}{%
\vrt{0}{0}
\dotted
\arrPT{2.4 0}{2.5 0}
\linePT{0}{0}{4.}{0}{}
\subp{4.}{0}{%
\txt{-0.5}{0.5}{\makebox[0ex][l]{$t$}}
}
\arrPT{-1.13137 1.13137}{-1.06066 1.06066}
\linePT{0}{0}{-2.82843}{2.82843}{}
\subp{-2.82843}{2.82843}{%
\txt{-0.5}{0}{\makebox[0ex][r]{$t'$}}
}
\arrPT{-1.6 0}{-1.5 0}
\linePT{0}{0}{-4.}{0}{}
\subp{-4.}{0}{%
\txt{0.5}{0.5}{\makebox[0ex][r]{$t''$}}
}
\arrPT{-1.69706 -1.69706}{-1.76777 -1.76777}
\linePT{0}{0}{-2.82843}{-2.82843}{}
\subp{-2.82843}{-2.82843}{%
\txt{-0.5}{0}{\makebox[0ex][r]{$t'''$}}
}
\tarr{-2.66421}{-4.82843}{4.}
}
}
}
\showPT{\figPTch}{figPTch}
\newcommand{\figPTci}{%
\dframe{10.3284}{6.07843}{\subp{5.82843}{0.25}{%
\vrt{0}{0}
\dotted
\arrPT{1.6 0}{1.5 0}
\linePT{0}{0}{4.}{0}{}
\subp{4.}{0}{%
\txt{-0.5}{0.5}{\makebox[0ex][l]{$t$}}
}
\arrPT{-1.69706 1.69706}{-1.76777 1.76777}
\linePT{0}{0}{-2.82843}{2.82843}{}
\subp{-2.82843}{2.82843}{%
\txt{-0.5}{0}{\makebox[0ex][r]{$t'$}}
}
\arrPT{-2.4 0}{-2.5 0}
\linePT{0}{0}{-4.}{0}{}
\subp{-4.}{0}{%
\txt{0.5}{0.5}{\makebox[0ex][r]{$t''$}}
}
\arrPT{-1.13137 -1.13137}{-1.06066 -1.06066}
\linePT{0}{0}{-2.82843}{-2.82843}{}
\subp{-2.82843}{-2.82843}{%
\txt{-0.5}{0}{\makebox[0ex][r]{$t'''$}}
}
\tarr{-2.66421}{-4.82843}{4.}
}
}
}
\showPT{\figPTci}{figPTci}
\newcommand{\figPTcj}{%
\dframe{10.6569}{6.07843}{\subp{5.82843}{0.25}{%
\vrt{0}{0}
\dotted
\arrPT{1.69706 1.69706}{1.76777 1.76777}
\linePT{0}{0}{2.82843}{2.82843}{}
\subp{2.82843}{2.82843}{%
\txt{0.5}{0}{\makebox[0ex][l]{$t$}}
}
\arrPT{1.69706 -1.69706}{1.76777 -1.76777}
\linePT{0}{0}{2.82843}{-2.82843}{}
\subp{2.82843}{-2.82843}{%
\txt{0.5}{0.5}{\makebox[0ex][l]{$t'$}}
}
\arrPT{-1.13137 1.13137}{-1.06066 1.06066}
\linePT{0}{0}{-2.82843}{2.82843}{}
\subp{-2.82843}{2.82843}{%
\txt{-0.5}{0}{\makebox[0ex][r]{$t''$}}
}
\arrPT{-1.13137 -1.13137}{-1.06066 -1.06066}
\linePT{0}{0}{-2.82843}{-2.82843}{}
\subp{-2.82843}{-2.82843}{%
\txt{-0.5}{0}{\makebox[0ex][r]{$t'''$}}
}
\tarr{-2.5}{-4.82843}{4.}
}
}
}
\showPT{\figPTcj}{figPTcj}
\newcommand{\figPTck}{%
\dframe{10.6569}{6.07843}{\subp{5.82843}{0.25}{%
\vrt{0}{0}
\dotted
\arrPT{1.13137 1.13137}{1.06066 1.06066}
\linePT{0}{0}{2.82843}{2.82843}{}
\subp{2.82843}{2.82843}{%
\txt{0.5}{0}{\makebox[0ex][l]{$t$}}
}
\arrPT{1.13137 -1.13137}{1.06066 -1.06066}
\linePT{0}{0}{2.82843}{-2.82843}{}
\subp{2.82843}{-2.82843}{%
\txt{0.5}{0.5}{\makebox[0ex][l]{$t'$}}
}
\arrPT{-1.69706 1.69706}{-1.76777 1.76777}
\linePT{0}{0}{-2.82843}{2.82843}{}
\subp{-2.82843}{2.82843}{%
\txt{-0.5}{0}{\makebox[0ex][r]{$t''$}}
}
\arrPT{-1.69706 -1.69706}{-1.76777 -1.76777}
\linePT{0}{0}{-2.82843}{-2.82843}{}
\subp{-2.82843}{-2.82843}{%
\txt{-0.5}{0}{\makebox[0ex][r]{$t'''$}}
}
\tarr{-2.5}{-4.82843}{4.}
}
}
}
\showPT{\figPTck}{figPTck}
\newcommand{\figPTcl}{%
\dframe{25.}{5.06891}{\subp{0.5}{1.25951}{%
\vrt{0}{0}
\solid
\arrPT{4.4 0}{4.5 0}
\linePT{0}{0}{8.}{0}{}
\subp{8.}{0}{%
\vrt{0}{0}
\arrPT{-1.13137 -1.13137}{-1.06066 -1.06066}
\linePT{0}{0}{-2.82843}{-2.82843}{}
\arrPT{4.4 0}{4.5 0}
\linePT{0}{0}{8.}{0}{}
\arrPT{4.45 2.28226}{4.5 2.28226}
\arcPT{-120.}{0 0}{4. -2.3094}
\subp{8.}{0}{%
\vrt{0}{0}
\arrPT{-1.69706 -1.69706}{-1.76777 -1.76777}
\linePT{0}{0}{-2.82843}{-2.82843}{}
\arrPT{4.4 0}{4.5 0}
\linePT{0}{0}{8.}{0}{}
\subp{8.}{0}{%
\vrt{0}{0}
}
}
}
\tarr{10.}{-4.82843}{4.}
}
}
}
\showPT{\figPTcl}{figPTcl}
\newcommand{\figPTcm}{%
\dframe{6.}{3.5}{\subp{1.5}{0.5}{%
\dotted
\txt{0.5}{0.5}{\makebox[0ex][r]{$r,t$}}
\arrPT{2.4 0}{2.5 0}
\linePT{0}{0}{4.}{0}{}
\subp{4.}{0}{%
\vrt{0}{0}
}
\tarr{-0.5}{-2.5}{4.}
}
}
}
\showPT{\figPTcm}{figPTcm}
\newcommand{\figPTcn}{%
\dframe{6.}{3.5}{\subp{1.5}{0.5}{%
\dotted
\txt{0.5}{0.5}{\makebox[0ex][r]{$r,t$}}
\arrPT{1.6 0}{1.5 0}
\linePT{0}{0}{4.}{0}{}
\subp{4.}{0}{%
\vrt{0}{0}
}
\tarr{-0.5}{-2.5}{4.}
}
}
}
\showPT{\figPTcn}{figPTcn}
\newcommand{\figPTco}{%
\dframe{6.}{3.5}{\subp{0.5}{0.5}{%
\vrt{0}{0}
\dotted
\arrPT{2.4 0}{2.5 0}
\linePT{0}{0}{4.}{0}{}
\subp{4.}{0}{%
\txt{-0.5}{0.5}{\makebox[0ex][l]{$r,t$}}
}
\tarr{0.5}{-2.5}{4.}
}
}
}
\showPT{\figPTco}{figPTco}
\newcommand{\figPTcp}{%
\dframe{6.}{3.5}{\subp{0.5}{0.5}{%
\vrt{0}{0}
\dotted
\arrPT{1.6 0}{1.5 0}
\linePT{0}{0}{4.}{0}{}
\subp{4.}{0}{%
\txt{-0.5}{0.5}{\makebox[0ex][l]{$r,t$}}
}
\tarr{0.5}{-2.5}{4.}
}
}
}
\showPT{\figPTcp}{figPTcp}
\newcommand{\figPTcq}{%
\dframe{5.5}{3.25}{\subp{0}{0.25}{%
\arrPT{2.4 0}{2.5 0}
\linePT{0}{0}{4.}{0}{}
\subp{4.}{0}{%
\txt{-0.5}{0.5}{\makebox[0ex][l]{$r,t$}}
}
\tarr{0.75}{-2.}{4.}
}
}
}
\showPT{\figPTcq}{figPTcq}
\newcommand{\figPTcr}{%
\dframe{5.5}{3.25}{\subp{0}{0.25}{%
\arrPT{1.6 0}{1.5 0}
\linePT{0}{0}{4.}{0}{}
\subp{4.}{0}{%
\txt{-0.5}{0.5}{\makebox[0ex][l]{$r,t$}}
}
\tarr{0.75}{-2.}{4.}
}
}
}
\showPT{\figPTcr}{figPTcr}
\newcommand{\figPTcs}{%
\dframe{9.5}{3.5}{\subp{1.}{0.}{%
\txt{0.5}{0.5}{\makebox[0ex][r]{$t'$}}
\arrPT{4.4 0}{4.5 0}
\linePT{0}{0}{8.}{0}{}
\subp{8.}{0}{%
\txt{-0.5}{0.5}{\makebox[0ex][l]{$t$}}
}
\txt{3.75}{1.}{\makebox[0ex][l]{$r$}}
\tarr{1.75}{-2.}{4.}
}
}
}
\showPT{\figPTcs}{figPTcs}
\newcommand{\figPTct}{%
\dframe{9.5}{3.5}{\subp{1.}{0.}{%
\txt{0.5}{0.5}{\makebox[0ex][r]{$t'$}}
\txt{3.75}{1.}{\makebox[0ex][l]{$r$}}
\arrPT{3.6 0}{3.5 0}
\linePT{0}{0}{8.}{0}{}
\subp{8.}{0}{%
\txt{-0.5}{0.5}{\makebox[0ex][l]{$t$}}
}
\tarr{1.75}{-2.}{4.}
}
}
}
\showPT{\figPTct}{figPTct}
\newcommand{\figPTcu}{%
\dframe{12.3284}{6.07843}{\subp{6.82843}{0.25}{%
\vrt{0}{0}
\dotted
\arrPT{2.4 0}{2.5 0}
\linePT{0}{0}{4.}{0}{}
\subp{4.}{0}{%
\txt{-0.5}{0.5}{\makebox[0ex][l]{$1,t$}}
}
\arrPT{-1.13137 1.13137}{-1.06066 1.06066}
\linePT{0}{0}{-2.82843}{2.82843}{}
\subp{-2.82843}{2.82843}{%
\txt{-0.5}{0}{\makebox[0ex][r]{$1,t'$}}
}
\arrPT{-1.6 0}{-1.5 0}
\linePT{0}{0}{-4.}{0}{}
\subp{-4.}{0}{%
\txt{0.5}{0.5}{\makebox[0ex][r]{$1,t''$}}
}
\arrPT{-1.69706 -1.69706}{-1.76777 -1.76777}
\linePT{0}{0}{-2.82843}{-2.82843}{}
\subp{-2.82843}{-2.82843}{%
\txt{-0.5}{0}{\makebox[0ex][r]{$2,t'''$}}
}
\tarr{-2.66421}{-4.82843}{4.}
}
}
}
\showPT{\figPTcu}{figPTcu}
\newcommand{\figPTcv}{%
\dframe{12.3284}{6.07843}{\subp{6.82843}{0.25}{%
\vrt{0}{0}
\dotted
\arrPT{1.6 0}{1.5 0}
\linePT{0}{0}{4.}{0}{}
\subp{4.}{0}{%
\txt{-0.5}{0.5}{\makebox[0ex][l]{$1,t$}}
}
\arrPT{-1.69706 1.69706}{-1.76777 1.76777}
\linePT{0}{0}{-2.82843}{2.82843}{}
\subp{-2.82843}{2.82843}{%
\txt{-0.5}{0}{\makebox[0ex][r]{$1,t'$}}
}
\arrPT{-2.4 0}{-2.5 0}
\linePT{0}{0}{-4.}{0}{}
\subp{-4.}{0}{%
\txt{0.5}{0.5}{\makebox[0ex][r]{$1,t''$}}
}
\arrPT{-1.13137 -1.13137}{-1.06066 -1.06066}
\linePT{0}{0}{-2.82843}{-2.82843}{}
\subp{-2.82843}{-2.82843}{%
\txt{-0.5}{0}{\makebox[0ex][r]{$2,t'''$}}
}
\tarr{-2.66421}{-4.82843}{4.}
}
}
}
\showPT{\figPTcv}{figPTcv}
\newcommand{\figPTcw}{%
\dframe{12.3284}{6.07843}{\subp{6.82843}{0.25}{%
\vrt{0}{0}
\dotted
\arrPT{2.4 0}{2.5 0}
\linePT{0}{0}{4.}{0}{}
\subp{4.}{0}{%
\txt{-0.5}{0.5}{\makebox[0ex][l]{$2,t$}}
}
\arrPT{-1.13137 1.13137}{-1.06066 1.06066}
\linePT{0}{0}{-2.82843}{2.82843}{}
\subp{-2.82843}{2.82843}{%
\txt{-0.5}{0}{\makebox[0ex][r]{$2,t'$}}
}
\arrPT{-1.6 0}{-1.5 0}
\linePT{0}{0}{-4.}{0}{}
\subp{-4.}{0}{%
\txt{0.5}{0.5}{\makebox[0ex][r]{$2,t''$}}
}
\arrPT{-1.69706 -1.69706}{-1.76777 -1.76777}
\linePT{0}{0}{-2.82843}{-2.82843}{}
\subp{-2.82843}{-2.82843}{%
\txt{-0.5}{0}{\makebox[0ex][r]{$1,t'''$}}
}
\tarr{-2.66421}{-4.82843}{4.}
}
}
}
\showPT{\figPTcw}{figPTcw}
\newcommand{\figPTcx}{%
\dframe{12.3284}{6.07843}{\subp{6.82843}{0.25}{%
\vrt{0}{0}
\dotted
\arrPT{1.6 0}{1.5 0}
\linePT{0}{0}{4.}{0}{}
\subp{4.}{0}{%
\txt{-0.5}{0.5}{\makebox[0ex][l]{$2,t$}}
}
\arrPT{-1.69706 1.69706}{-1.76777 1.76777}
\linePT{0}{0}{-2.82843}{2.82843}{}
\subp{-2.82843}{2.82843}{%
\txt{-0.5}{0}{\makebox[0ex][r]{$2,t'$}}
}
\arrPT{-2.4 0}{-2.5 0}
\linePT{0}{0}{-4.}{0}{}
\subp{-4.}{0}{%
\txt{0.5}{0.5}{\makebox[0ex][r]{$2,t''$}}
}
\arrPT{-1.13137 -1.13137}{-1.06066 -1.06066}
\linePT{0}{0}{-2.82843}{-2.82843}{}
\subp{-2.82843}{-2.82843}{%
\txt{-0.5}{0}{\makebox[0ex][r]{$1,t'''$}}
}
\tarr{-2.66421}{-4.82843}{4.}
}
}
}
\showPT{\figPTcx}{figPTcx}
\newcommand{\figPTcy}{%
\dframe{12.6569}{6.07843}{\subp{6.82843}{0.25}{%
\vrt{0}{0}
\dotted
\arrPT{1.69706 1.69706}{1.76777 1.76777}
\linePT{0}{0}{2.82843}{2.82843}{}
\subp{2.82843}{2.82843}{%
\txt{0.5}{0}{\makebox[0ex][l]{$1,t$}}
}
\arrPT{1.69706 -1.69706}{1.76777 -1.76777}
\linePT{0}{0}{2.82843}{-2.82843}{}
\subp{2.82843}{-2.82843}{%
\txt{0.5}{0.5}{\makebox[0ex][l]{$1,t'$}}
}
\arrPT{-1.13137 1.13137}{-1.06066 1.06066}
\linePT{0}{0}{-2.82843}{2.82843}{}
\subp{-2.82843}{2.82843}{%
\txt{-0.5}{0}{\makebox[0ex][r]{$1,t''$}}
}
\arrPT{-1.13137 -1.13137}{-1.06066 -1.06066}
\linePT{0}{0}{-2.82843}{-2.82843}{}
\subp{-2.82843}{-2.82843}{%
\txt{-0.5}{0}{\makebox[0ex][r]{$1,t'''$}}
}
\tarr{-2.5}{-4.82843}{4.}
}
}
}
\showPT{\figPTcy}{figPTcy}
\newcommand{\figPTcz}{%
\dframe{12.6569}{6.07843}{\subp{6.82843}{0.25}{%
\vrt{0}{0}
\dotted
\arrPT{1.13137 1.13137}{1.06066 1.06066}
\linePT{0}{0}{2.82843}{2.82843}{}
\subp{2.82843}{2.82843}{%
\txt{0.5}{0}{\makebox[0ex][l]{$1,t$}}
}
\arrPT{1.13137 -1.13137}{1.06066 -1.06066}
\linePT{0}{0}{2.82843}{-2.82843}{}
\subp{2.82843}{-2.82843}{%
\txt{0.5}{0.5}{\makebox[0ex][l]{$1,t'$}}
}
\arrPT{-1.69706 1.69706}{-1.76777 1.76777}
\linePT{0}{0}{-2.82843}{2.82843}{}
\subp{-2.82843}{2.82843}{%
\txt{-0.5}{0}{\makebox[0ex][r]{$1,t''$}}
}
\arrPT{-1.69706 -1.69706}{-1.76777 -1.76777}
\linePT{0}{0}{-2.82843}{-2.82843}{}
\subp{-2.82843}{-2.82843}{%
\txt{-0.5}{0}{\makebox[0ex][r]{$1,t'''$}}
}
\tarr{-2.5}{-4.82843}{4.}
}
}
}
\showPT{\figPTcz}{figPTcz}

\newcommand{\figPTda}{%
\dframe{12.6569}{6.07843}{\subp{6.82843}{0.25}{%
\vrt{0}{0}
\dotted
\arrPT{1.69706 1.69706}{1.76777 1.76777}
\linePT{0}{0}{2.82843}{2.82843}{}
\subp{2.82843}{2.82843}{%
\txt{0.5}{0}{\makebox[0ex][l]{$2,t$}}
}
\arrPT{1.69706 -1.69706}{1.76777 -1.76777}
\linePT{0}{0}{2.82843}{-2.82843}{}
\subp{2.82843}{-2.82843}{%
\txt{0.5}{0.5}{\makebox[0ex][l]{$2,t'$}}
}
\arrPT{-1.13137 1.13137}{-1.06066 1.06066}
\linePT{0}{0}{-2.82843}{2.82843}{}
\subp{-2.82843}{2.82843}{%
\txt{-0.5}{0}{\makebox[0ex][r]{$2,t''$}}
}
\arrPT{-1.13137 -1.13137}{-1.06066 -1.06066}
\linePT{0}{0}{-2.82843}{-2.82843}{}
\subp{-2.82843}{-2.82843}{%
\txt{-0.5}{0}{\makebox[0ex][r]{$2,t'''$}}
}
\tarr{-2.5}{-4.82843}{4.}
}
}
}
\showPT{\figPTda}{figPTda}
\newcommand{\figPTdb}{%
\dframe{12.6569}{6.07843}{\subp{6.82843}{0.25}{%
\vrt{0}{0}
\dotted
\arrPT{1.13137 1.13137}{1.06066 1.06066}
\linePT{0}{0}{2.82843}{2.82843}{}
\subp{2.82843}{2.82843}{%
\txt{0.5}{0}{\makebox[0ex][l]{$2,t$}}
}
\arrPT{1.13137 -1.13137}{1.06066 -1.06066}
\linePT{0}{0}{2.82843}{-2.82843}{}
\subp{2.82843}{-2.82843}{%
\txt{0.5}{0.5}{\makebox[0ex][l]{$2,t'$}}
}
\arrPT{-1.69706 1.69706}{-1.76777 1.76777}
\linePT{0}{0}{-2.82843}{2.82843}{}
\subp{-2.82843}{2.82843}{%
\txt{-0.5}{0}{\makebox[0ex][r]{$2,t''$}}
}
\arrPT{-1.69706 -1.69706}{-1.76777 -1.76777}
\linePT{0}{0}{-2.82843}{-2.82843}{}
\subp{-2.82843}{-2.82843}{%
\txt{-0.5}{0}{\makebox[0ex][r]{$2,t'''$}}
}
\tarr{-2.5}{-4.82843}{4.}
}
}
}
\showPT{\figPTdb}{figPTdb}
\newcommand{\figPTdc}{%
\dframe{12.6569}{6.07843}{\subp{6.82843}{0.25}{%
\vrt{0}{0}
\dotted
\arrPT{1.69706 1.69706}{1.76777 1.76777}
\linePT{0}{0}{2.82843}{2.82843}{}
\subp{2.82843}{2.82843}{%
\txt{0.5}{0}{\makebox[0ex][l]{$1,t$}}
}
\arrPT{1.13137 -1.13137}{1.06066 -1.06066}
\linePT{0}{0}{2.82843}{-2.82843}{}
\subp{2.82843}{-2.82843}{%
\txt{0.5}{0.5}{\makebox[0ex][l]{$1,t'$}}
}
\arrPT{-1.13137 1.13137}{-1.06066 1.06066}
\linePT{0}{0}{-2.82843}{2.82843}{}
\subp{-2.82843}{2.82843}{%
\txt{-0.5}{0}{\makebox[0ex][r]{$1,t''$}}
}
\arrPT{-1.69706 -1.69706}{-1.76777 -1.76777}
\linePT{0}{0}{-2.82843}{-2.82843}{}
\subp{-2.82843}{-2.82843}{%
\txt{-0.5}{0}{\makebox[0ex][r]{$1,t'''$}}
}
\tarr{-2.5}{-4.82843}{4.}
}
}
}
\showPT{\figPTdc}{figPTdc}
\newcommand{\figPTdd}{%
\dframe{12.6569}{6.07843}{\subp{6.82843}{0.25}{%
\vrt{0}{0}
\dotted
\arrPT{1.69706 1.69706}{1.76777 1.76777}
\linePT{0}{0}{2.82843}{2.82843}{}
\subp{2.82843}{2.82843}{%
\txt{0.5}{0}{\makebox[0ex][l]{$2,t$}}
}
\arrPT{1.13137 -1.13137}{1.06066 -1.06066}
\linePT{0}{0}{2.82843}{-2.82843}{}
\subp{2.82843}{-2.82843}{%
\txt{0.5}{0.5}{\makebox[0ex][l]{$2,t'$}}
}
\arrPT{-1.13137 1.13137}{-1.06066 1.06066}
\linePT{0}{0}{-2.82843}{2.82843}{}
\subp{-2.82843}{2.82843}{%
\txt{-0.5}{0}{\makebox[0ex][r]{$2,t''$}}
}
\arrPT{-1.69706 -1.69706}{-1.76777 -1.76777}
\linePT{0}{0}{-2.82843}{-2.82843}{}
\subp{-2.82843}{-2.82843}{%
\txt{-0.5}{0}{\makebox[0ex][r]{$2,t'''$}}
}
\tarr{-2.5}{-4.82843}{4.}
}
}
}
\showPT{\figPTdd}{figPTdd}
\newcommand{\figPTde}{%
\dframe{12.6569}{6.07843}{\subp{6.82843}{0.25}{%
\vrt{0}{0}
\dotted
\arrPT{1.69706 1.69706}{1.76777 1.76777}
\linePT{0}{0}{2.82843}{2.82843}{}
\subp{2.82843}{2.82843}{%
\txt{0.5}{0}{\makebox[0ex][l]{$1,t$}}
}
\arrPT{1.69706 -1.69706}{1.76777 -1.76777}
\linePT{0}{0}{2.82843}{-2.82843}{}
\subp{2.82843}{-2.82843}{%
\txt{0.5}{0.5}{\makebox[0ex][l]{$2,t'$}}
}
\arrPT{-1.13137 1.13137}{-1.06066 1.06066}
\linePT{0}{0}{-2.82843}{2.82843}{}
\subp{-2.82843}{2.82843}{%
\txt{-0.5}{0}{\makebox[0ex][r]{$1,t''$}}
}
\arrPT{-1.13137 -1.13137}{-1.06066 -1.06066}
\linePT{0}{0}{-2.82843}{-2.82843}{}
\subp{-2.82843}{-2.82843}{%
\txt{-0.5}{0}{\makebox[0ex][r]{$2,t'''$}}
}
\tarr{-2.5}{-4.82843}{4.}
}
}
}
\showPT{\figPTde}{figPTde}
\newcommand{\figPTdf}{%
\dframe{12.6569}{6.07843}{\subp{6.82843}{0.25}{%
\vrt{0}{0}
\dotted
\arrPT{1.13137 1.13137}{1.06066 1.06066}
\linePT{0}{0}{2.82843}{2.82843}{}
\subp{2.82843}{2.82843}{%
\txt{0.5}{0}{\makebox[0ex][l]{$1,t$}}
}
\arrPT{1.13137 -1.13137}{1.06066 -1.06066}
\linePT{0}{0}{2.82843}{-2.82843}{}
\subp{2.82843}{-2.82843}{%
\txt{0.5}{0.5}{\makebox[0ex][l]{$2,t'$}}
}
\arrPT{-1.69706 1.69706}{-1.76777 1.76777}
\linePT{0}{0}{-2.82843}{2.82843}{}
\subp{-2.82843}{2.82843}{%
\txt{-0.5}{0}{\makebox[0ex][r]{$1,t''$}}
}
\arrPT{-1.69706 -1.69706}{-1.76777 -1.76777}
\linePT{0}{0}{-2.82843}{-2.82843}{}
\subp{-2.82843}{-2.82843}{%
\txt{-0.5}{0}{\makebox[0ex][r]{$2,t'''$}}
}
\tarr{-2.5}{-4.82843}{4.}
}
}
}
\showPT{\figPTdf}{figPTdf}
\newcommand{\figPTdg}{%
\dframe{12.6569}{6.07843}{\subp{6.82843}{0.25}{%
\vrt{0}{0}
\dotted
\arrPT{1.69706 1.69706}{1.76777 1.76777}
\linePT{0}{0}{2.82843}{2.82843}{}
\subp{2.82843}{2.82843}{%
\txt{0.5}{0}{\makebox[0ex][l]{$1,t$}}
}
\arrPT{1.13137 -1.13137}{1.06066 -1.06066}
\linePT{0}{0}{2.82843}{-2.82843}{}
\subp{2.82843}{-2.82843}{%
\txt{0.5}{0.5}{\makebox[0ex][l]{$2,t'$}}
}
\arrPT{-1.13137 1.13137}{-1.06066 1.06066}
\linePT{0}{0}{-2.82843}{2.82843}{}
\subp{-2.82843}{2.82843}{%
\txt{-0.5}{0}{\makebox[0ex][r]{$1,t''$}}
}
\arrPT{-1.69706 -1.69706}{-1.76777 -1.76777}
\linePT{0}{0}{-2.82843}{-2.82843}{}
\subp{-2.82843}{-2.82843}{%
\txt{-0.5}{0}{\makebox[0ex][r]{$2,t'''$}}
}
\tarr{-2.5}{-4.82843}{4.}
}
}
}
\showPT{\figPTdg}{figPTdg}
\newcommand{\figPTdh}{%
\dframe{12.6569}{6.07843}{\subp{6.82843}{0.25}{%
\vrt{0}{0}
\dotted
\arrPT{1.13137 1.13137}{1.06066 1.06066}
\linePT{0}{0}{2.82843}{2.82843}{}
\subp{2.82843}{2.82843}{%
\txt{0.5}{0}{\makebox[0ex][l]{$1,t$}}
}
\arrPT{1.69706 -1.69706}{1.76777 -1.76777}
\linePT{0}{0}{2.82843}{-2.82843}{}
\subp{2.82843}{-2.82843}{%
\txt{0.5}{0.5}{\makebox[0ex][l]{$2,t'$}}
}
\arrPT{-1.69706 1.69706}{-1.76777 1.76777}
\linePT{0}{0}{-2.82843}{2.82843}{}
\subp{-2.82843}{2.82843}{%
\txt{-0.5}{0}{\makebox[0ex][r]{$1,t''$}}
}
\arrPT{-1.13137 -1.13137}{-1.06066 -1.06066}
\linePT{0}{0}{-2.82843}{-2.82843}{}
\subp{-2.82843}{-2.82843}{%
\txt{-0.5}{0}{\makebox[0ex][r]{$2,t'''$}}
}
\tarr{-2.5}{-4.82843}{4.}
}
}
}
\showPT{\figPTdh}{figPTdh}
\newcommand{\figPTdi}{%
\dframe{17.}{8.65685}{\subp{8.5}{1.}{%
\vrt{0}{0}
\arrPT{4.4 0}{4.5 0}
\linePT{0}{0}{8.}{0}{}
\subp{8.}{0}{%
\vrt{0}{0}
}
\txt{4.5}{1.}{\makebox[0ex][l]{$1$}}
\arrPT{-3.6 0}{-3.5 0}
\linePT{0}{0}{-8.}{0}{}
\subp{-8.}{0}{%
\vrt{0}{0}
}
\txt{-5.5}{1.}{\makebox[0ex][l]{$1$}}
\arrPT{-2.54558 2.54558}{-2.47487 2.47487}
\linePT{0}{0}{-5.65685}{5.65685}{}
\subp{-5.65685}{5.65685}{%
\vrt{0}{0}
}
\txt{-4.03553}{4.53553}{\makebox[0ex][l]{$1$}}
\arrPT{-3.11127 -3.11127}{-3.18198 -3.18198}
\linePT{0}{0}{-5.65685}{-5.65685}{}
\subp{-5.65685}{-5.65685}{%
\vrt{0}{0}
}
\txt{-4.03553}{-2.53553}{\makebox[0ex][l]{$2$}}
\tarr{-2.}{-8.15685}{4.}
}
}
}
\showPT{\figPTdi}{figPTdi}
\newcommand{\figPTdj}{%
\dframe{17.}{8.65685}{\subp{8.5}{1.}{%
\vrt{0}{0}
\arrPT{4.4 0}{4.5 0}
\linePT{0}{0}{8.}{0}{}
\subp{8.}{0}{%
\vrt{0}{0}
}
\txt{4.5}{1.}{\makebox[0ex][l]{$2$}}
\arrPT{-3.6 0}{-3.5 0}
\linePT{0}{0}{-8.}{0}{}
\subp{-8.}{0}{%
\vrt{0}{0}
}
\txt{-5.5}{1.}{\makebox[0ex][l]{$2$}}
\arrPT{-2.54558 2.54558}{-2.47487 2.47487}
\linePT{0}{0}{-5.65685}{5.65685}{}
\subp{-5.65685}{5.65685}{%
\vrt{0}{0}
}
\txt{-4.03553}{4.53553}{\makebox[0ex][l]{$2$}}
\arrPT{-3.11127 -3.11127}{-3.18198 -3.18198}
\linePT{0}{0}{-5.65685}{-5.65685}{}
\subp{-5.65685}{-5.65685}{%
\vrt{0}{0}
}
\txt{-4.03553}{-2.53553}{\makebox[0ex][l]{$1$}}
\tarr{-2.}{-8.15685}{4.}
}
}
}
\showPT{\figPTdj}{figPTdj}
\newcommand{\figPTdk}{%
\dframe{5.5}{3.25}{\subp{0}{0.25}{%
\arrPT{2.4 0}{2.5 0}
\linePT{0}{0}{4.}{0}{}
\subp{4.}{0}{%
\txt{-0.5}{0.5}{\makebox[0ex][l]{$1,t$}}
}
\tarr{0.75}{-2.}{4.}
}
}
}
\showPT{\figPTdk}{figPTdk}
\newcommand{\figPTdl}{%
\dframe{5.5}{3.25}{\subp{0}{0.25}{%
\arrPT{1.6 0}{1.5 0}
\linePT{0}{0}{4.}{0}{}
\subp{4.}{0}{%
\txt{-0.5}{0.5}{\makebox[0ex][l]{$2,t$}}
}
\tarr{0.75}{-2.}{4.}
}
}
}
\showPT{\figPTdl}{figPTdl}
\newcommand{\figPTdm}{%
\dframe{9.5}{3.5}{\subp{1.}{0.}{%
\txt{0.5}{0.5}{\makebox[0ex][r]{$t'$}}
\arrPT{4.4 0}{4.5 0}
\linePT{0}{0}{8.}{0}{}
\subp{8.}{0}{%
\txt{-0.5}{0.5}{\makebox[0ex][l]{$t$}}
}
\txt{3.75}{1.}{\makebox[0ex][l]{$1$}}
\tarr{1.75}{-2.}{4.}
}
}
}
\showPT{\figPTdm}{figPTdm}
\newcommand{\figPTdn}{%
\dframe{9.5}{3.5}{\subp{1.}{0.}{%
\txt{0.5}{0.5}{\makebox[0ex][r]{$t'$}}
\txt{3.75}{1.}{\makebox[0ex][l]{$2$}}
\arrPT{3.6 0}{3.5 0}
\linePT{0}{0}{8.}{0}{}
\subp{8.}{0}{%
\txt{-0.5}{0.5}{\makebox[0ex][l]{$t$}}
}
\tarr{1.75}{-2.}{4.}
}
}
}
\showPT{\figPTdn}{figPTdn}
\newcommand{\figPTdo}{%
\dframe{9.}{4.}{\subp{0.5}{0.}{%
\solid
\vrt{0}{0}
\linePT{0}{0}{8.}{0}{}
\subp{8.}{0}{%
\vrt{0}{0}
}
\vrt{0}{2.}
\linePT{0}{2.}{8.}{2.}{}
\subp{8.}{2.}{%
\vrt{0}{0}
}
\tarr{2.}{-2.5}{4.}
}
}
}
\showPT{\figPTdo}{figPTdo}
\newcommand{\figPTdp}{%
\dframe{9.}{5.}{\subp{0.5}{-1.}{%
\solid
\vrt{0}{0}
\linePT{0}{0}{8.}{0}{}
\subp{8.}{0}{%
\vrt{0}{0}
}
\vrt{0}{2.}
\linePT{0}{2.}{8.}{2.}{}
\subp{8.}{2.}{%
\vrt{0}{0}
}
\vrt{0}{4.}
\linePT{0}{4.}{8.}{4.}{}
\subp{8.}{4.}{%
\vrt{0}{0}
}
\tarr{2.}{-2.5}{4.}
}
}
}
\showPT{\figPTdp}{figPTdp}
\newcommand{\figPTdq}{%
\dframe{9.}{6.}{\subp{0.5}{-2.}{%
\solid
\vrt{0}{0}
\linePT{0}{0}{8.}{0}{}
\subp{8.}{0}{%
\vrt{0}{0}
}
\vrt{0}{2.}
\linePT{0}{2.}{8.}{2.}{}
\subp{8.}{2.}{%
\vrt{0}{0}
}
\vrt{0}{4.}
\linePT{0}{4.}{8.}{4.}{}
\subp{8.}{4.}{%
\vrt{0}{0}
}
\vrt{0}{6.}
\linePT{0}{6.}{8.}{6.}{}
\subp{8.}{6.}{%
\vrt{0}{0}
}
\tarr{2.}{-2.5}{4.}
}
}
}
\showPT{\figPTdq}{figPTdq}
\newcommand{\figPTdr}{%
\dframe{9.}{8.}{\subp{0.5}{0.}{%
\solid
\vrt{0}{2.}
\linePT{0}{2.}{6.9282}{6.}{}
\subp{6.9282}{6.}{%
\vrt{0}{0}
}
\linePT{0}{2.}{8.}{2.}{}
\subp{8.}{2.}{%
\vrt{0}{0}
}
\vrt{0}{0}
\linePT{0}{0}{8.}{0}{}
\subp{8.}{0}{%
\vrt{0}{0}
}
\linePT{0}{0}{6.9282}{-4.}{}
\subp{6.9282}{-4.}{%
\vrt{0}{0}
}
\tarr{2.}{-6.5}{4.}
}
}
}
\showPT{\figPTdr}{figPTdr}
\newcommand{\figPTds}{%
\dframe{5.65685}{5.32843}{\subp{2.82843}{1.}{%
\solid\setdashpattern <1.ex, 0.5ex>
\vrt{0}{0}
\linePT{0}{0}{2.82843}{2.82843}{}
\linePT{0}{0}{-2.82843}{2.82843}{}
\linePT{0}{0}{2.82843}{-2.82843}{}
\linePT{0}{0}{-2.82843}{-2.82843}{}
\tarr{-2.}{-4.82843}{4.}
}
}
}
\showPT{\figPTds}{figPTds}
\newcommand{\figPTdt}{%
\dframe{12.5}{4.}{\subp{0}{0.}{%
\solid
\linePT{0}{0}{12.}{0}{}
\subp{12.}{0}{%
\vrt{0}{0}
\vrt{-8.}{0}
}
\linePT{0}{2.}{12.}{2.}{}
\subp{12.}{2.}{%
\vrt{0}{0}
\vrt{-8.}{0}
}
\tarr{4.25}{-2.5}{4.}
}
}
}
\showPT{\figPTdt}{figPTdt}
\newcommand{\figPTdu}{%
\dframe{10.8923}{7.}{\subp{3.4641}{1.}{%
\solid
\vrt{0}{0}
\linePT{0}{0}{6.9282}{4.}{}
\subp{6.9282}{4.}{%
\vrt{0}{0}
}
\linePT{0}{0}{6.9282}{-4.}{}
\subp{6.9282}{-4.}{%
\vrt{0}{0}
}
\linePT{0}{0}{-3.4641}{2.}{}
\linePT{0}{0}{-3.4641}{-2.}{}
\tarr{-0.0179492}{-6.5}{4.}
}
}
}
\showPT{\figPTdu}{figPTdu}
\newcommand{\figPTdv}{%
\dframe{10.4069}{5.82843}{\subp{5.32843}{0.5}{%
\dotted
\vrt{0}{0}
\linePT{0}{0}{2.82843}{2.82843}{}
\subp{2.82843}{2.82843}{%
\txt{0.75}{0}{\makebox[0ex][l]{$x_1$}}
}
\linePT{0}{0}{-2.82843}{2.82843}{}
\subp{-2.82843}{2.82843}{%
\txt{-0.5}{0}{\makebox[0ex][r]{$x'_1$}}
}
\linePT{0}{0}{2.82843}{-2.82843}{}
\subp{2.82843}{-2.82843}{%
\txt{0.75}{0}{\makebox[0ex][l]{$x_2$}}
}
\linePT{0}{0}{-2.82843}{-2.82843}{}
\subp{-2.82843}{-2.82843}{%
\txt{-0.5}{0}{\makebox[0ex][r]{$x'_2$}}
}
\tarr{-2.125}{-4.82843}{4.}
}
}
}
\showPT{\figPTdv}{figPTdv}
\newcommand{\figPTdw}{%
\dframe{7.4641}{4.5}{\subp{3.4641}{1.}{%
\solid\setdashpattern <1.ex, 0.5ex>
\vrt{0}{0}
\linePT{0}{0}{4.}{0}{}
\linePT{0}{0}{-3.4641}{2.}{}
\linePT{0}{0}{-3.4641}{-2.}{}
\tarr{-1.73205}{-4.}{4.}
}
}
}
\showPT{\figPTdw}{figPTdw}
\newcommand{\figPTdx}{%
\dframe{7.22287}{5.9641}{\subp{3.75877}{1.}{%
\solid\setdashpattern <1.ex, 0.5ex>
\vrt{0}{0}
\linePT{0}{0}{3.4641}{2.}{}
\linePT{0}{0}{3.4641}{-2.}{}
\linePT{0}{0}{-2.}{3.4641}{}
\linePT{0}{0}{-2.}{-3.4641}{}
\linePT{0}{0}{-3.75877}{1.36808}{}
\linePT{0}{0}{-3.75877}{-1.36808}{}
\tarr{-2.14733}{-5.4641}{4.}
}
}
}
\showPT{\figPTdx}{figPTdx}
\newcommand{\figPTdy}{%
\dframe{17.}{4.16421}{\subp{0.5}{-0.164214}{%
\solid
\txt{-0.5}{0.5}{\makebox[0ex][l]{$x'$}}
\vrt{8.}{0}
\linePT{8.}{0}{5.17157}{2.82843}{}
\linePT{0}{0}{16.}{0}{}
\subp{16.}{0}{%
\txt{-0.5}{0.5}{\makebox[0ex][l]{$x$}}
}
\tarr{6.}{-2.5}{4.}
}
}
}
\showPT{\figPTdy}{figPTdy}
\newcommand{\figPTdz}{%
\dframe{25.}{4.32843}{\subp{0.5}{0.828427}{%
\solid
\txt{-0.5}{0.5}{\makebox[0ex][l]{$x'$}}
\vrt{8.}{0}
\vrt{16.}{0}
\arcPT{-90.}{8. 0}{12. -4.}
\subp{16.}{0}{}
\arcPT{90.}{8. 0}{12. 4.}
\subp{16.}{0}{}
\linePT{0}{0}{8.}{0}{}
\linePT{16.}{0}{24.}{0}{}
\subp{24.}{0}{%
\txt{-0.5}{0.5}{\makebox[0ex][l]{$x$}}
}
\tarr{10.}{-3.65685}{4.}
}
}
}
\showPT{\figPTdz}{figPTdz}
\newcommand{\figPTea}{%
\dframe{9.5}{5.}{\subp{0.5}{-1.5}{%
\solid
\linePT{4.}{3.}{8.}{3.}{}
\subp{8.}{3.}{%
\txt{-0.5}{0.5}{\makebox[0ex][l]{$x$}}
}
\txt{-0.5}{0.5}{\makebox[0ex][l]{$x''$}}
\linePT{0}{0}{8.}{0}{}
\subp{8.}{0}{%
\txt{-0.5}{0.5}{\makebox[0ex][l]{$x'$}}
}
\tarr{2.25}{-2.}{4.}
}
}
}
\showPT{\figPTea}{figPTea}
\newcommand{\figPTeb}{%
\dframe{9.5}{5.}{\subp{0.5}{-1.5}{%
\solid
\linePT{4.}{3.}{8.}{3.}{}
\subp{8.}{3.}{%
\txt{-0.5}{0.5}{\makebox[0ex][l]{$x'$}}
}
\txt{-0.5}{0.5}{\makebox[0ex][l]{$x''$}}
\linePT{0}{0}{8.}{0}{}
\subp{8.}{0}{%
\txt{-0.5}{0.5}{\makebox[0ex][l]{$x$}}
}
\tarr{2.25}{-2.}{4.}
}
}
}
\showPT{\figPTeb}{figPTeb}
\newcommand{\figPTec}{%
\dframe{17.4282}{7.}{\subp{8.5}{1.}{%
\solid
\vrt{0}{0}
\linePT{0}{0}{-8.}{0}{}
\subp{-8.}{0}{%
\txt{-0.5}{0.5}{\makebox[0ex][l]{$x''$}}
}
\linePT{0}{0}{6.9282}{4.}{}
\subp{6.9282}{4.}{%
\txt{0.5}{-0.5}{\makebox[0ex][l]{$x$}}
}
\linePT{0}{0}{6.9282}{-4.}{}
\subp{6.9282}{-4.}{%
\txt{0.5}{-0.5}{\makebox[0ex][l]{$x'$}}
}
\tarr{-1.7859}{-6.5}{4.}
}
}
}
\showPT{\figPTec}{figPTec}
\newcommand{\figPTed}{%
\dframe{18.4282}{7.}{\subp{9.4282}{1.}{%
\solid
\vrt{0}{0}
\linePT{0}{0}{8.}{0}{}
\subp{8.}{0}{%
\txt{-0.5}{0.5}{\makebox[0ex][l]{$x$}}
}
\linePT{0}{0}{-6.9282}{4.}{}
\subp{-6.9282}{4.}{%
\txt{-2.5}{-0.5}{\makebox[0ex][l]{$x'$}}
}
\linePT{0}{0}{-6.9282}{-4.}{}
\subp{-6.9282}{-4.}{%
\txt{-2.5}{-0.5}{\makebox[0ex][l]{$x''$}}
}
\tarr{-2.2141}{-6.5}{4.}
}
}
}
\showPT{\figPTed}{figPTed}

\newcommand{\dd}{2.82843}

\newcommand{\cdf}[1]{%
\newcounter{#1a}
\newcounter{#1b}
\newcounter{#1c}
\newcounter{#1d}
\newcounter{#1e}
}

\cdf{c}

\renewcommand{\showlabel}[1]{\label{#1}}
\begin{document} 
\tighten
\widetext
\title{Diagram expansions in classical 
stochastic field theory}
\author{L. I. Plimak, M. Fleischhauer$^*$, M. J. Collett and D. F. Walls}
\address{Dept. of Physics, University of Auckland, Private Bag
92019, Auckland, New Zealand}
\address{$^*$ Sektion Physik, Ludwig-Maximilians Universit\"at M\"unchen,
D-80333 M\"unchen, Germany}
\date{\today}
\maketitle
\begin{abstract}
A diagram approach to classical nonlinear stochastic field theory  
 is introduced.
This approach is intended to serve as a link between quantum 
and classical field theories, resulting in an independent 
constructive characterisation of the measure in Feynman path 
integrals  
in terms of stochastic differential equations for the paths.
\end{abstract}
\section{Introduction}
Diagrams are commonly 
associated with either 
Feynman's work on quantum 
electrodynamics, 
or with Matsubara and Keldysh techniques in statistical 
mechanics, so that  diagrams appear to be something to do with  
subtle problems in relativistic quantum field theory 
or the theory of phase transitions. 
This view is totally misleading.   
The concept of diagrams is very general and 
can be applied to virtually any problem in physics. 
By encoding details of a physical problem 
as expressions 
for the lines and vertices,  
a solution to this problem 
is readily produced
following general diagram rules. 

There exists an additional dimension to the diagram  
approach not yet fully appreciated, namely its ability to 
establish links between approaches 
based on inconsistent techniques, like quantum and 
classical field theories (QFT and CFT). 
The diagram approach to the QFT is common knowledge.
In this paper,  we develop a diagram approach 
to the theory of stochastic differential equations (SDEs), 
which is the formal ground of the CFT. 
The central formal result of the paper is 
a general relation which we call the {\em causal Wick's 
theorem\/}. 
It plays the same role in deriving diagram series 
for SDEs as Wick's theorem in QFT 
plays in deriving the Feynman (say) diagram 
series. 
We then find that any SDE is  formally solved 
by diagram series of a certain 
structure, 
which we call {\em causal diagram series\/}.  
In these series,  the propagator 
is determined by the linear part of the SDE, and 
vertices by the nonlinearity and noise sources. 
This result may be interesting in itself since it 
allows one to extend methods specific to diagram 
approaches (e.g., the Dyson equations) to SDEs.

The formal relation between diagram series 
in  the QFT and in the CFT will be investigated in 
papers to follow. 
Here we note only 
that the (pseudo-)stochastic measure 
on the solutions of an SDE is exactly the measure in 
certain Feynman path integrals: solutions to the SDE {\em are\/} 
the Feynman paths. 
This may result in better understanding of the path-integral 
approach and also in the possibility of characterising path 
integrals constructively, in well-defined mathematical terms, 
and,  indeed  
in a way independent of perturbation theory. 

From an even more fundamental perspective, 
we encounter a very important link between 
regularisations in the diagrams  and stochastic calculus in SDEs. 
Technically, we find that {\em regularisations\/} 
in a diagram series which make  
diagrams convergent affect  the 
respective SDE in such a way that it becomes 
mathematically 
defined within the normal calculus. 
We also find some indication that a {\em renormalisation\/} 
procedure, 
which formally consists of (i) regularisation and (ii) a 
limiting procedure removing this regularisation in a 
certain way, 
appears 
from the point of view of the respective SDE 
as a limiting procedure specifying  
the stochastic calculus. 
This link between renormalisation and stochastic calculus 
suggests more extensive
mathematical investigation, 
which may lead to better understanding of the  
formal grounds of quantum field theory.

\section{Classical stochastic self-action problem}
\subsection{Stochastic calculus and regularisations in SDEs} 
We consider 
a c-number field $ \psi (r,t)$, 
which satisfies a 
generic equation with  
a source $s(r,t)$, 
\begin{eqnarray} 
\showlabel{BasicEq}
{\cal L}\psi (r,t) = s(r,t) ,
\end{eqnarray}%
where ${\cal L}$ is a  
differential operator. 
The source 
$s$ is dependent on the field $ \psi $ and this dependence 
may include randomness; 
that is, 
in general (\ref{BasicEq}) is a stochastic differential 
equation for the field $ \psi (r,t)$.
To be specific, we assumed that $ \psi $ is 
a classical field in one-dimensional space; 
to consider other situations, e.g., 
a multi-mode field, or a field 
in three-dimensional space, one should simply replace 
in the relations below 
$\int dr$ by $\sum_r$ (i.e., $r$ is then a mode 
index), or $\int d^3 {\bf r}$, respectively. 
For a single-mode field, the variable $r$ and the 
summation should simply be dropped. 
We shall often resort to this last case for simplicity of 
examples. 

Formally solving (\ref{BasicEq}) turns it into 
an integral equation,   
(assuming $s(r,t) \rightarrow 0$ as $t \rightarrow - \infty$) 
\begin{eqnarray} 
\showlabel{BasicInt}
 \psi (r,t) = \int _{-\infty}^{\infty}dt' 
\int dr' G(r,r',t-t')s(r',t') +  
\psi _0 (r,t),
\end{eqnarray}%
where  $G(r,r',t-t')$
is the retarded Green's function of the equation (\ref{BasicEq}),  
\begin{eqnarray} 
\showlabel{GGen}
{\cal L}G(r,r',t-t') =   \delta (r-r')\delta (t-t') , \ \ \
G(r,r',t-t') = 0, t<t' , 
\end{eqnarray}%
(so that integration in (\ref{BasicInt}) is in 
fact from minus infinity to $t$),  
and $ \psi  _0$ is the $in$-field, i.e. 
$ \psi  =  \psi  _0$ before the source is on. 
It obeys the free version of the equation (\ref{BasicEq}),  
\begin{eqnarray} 
\showlabel{BasicFree}
{\cal L}  \psi  _0 (r,t) = 0 .
\end{eqnarray}%
Note that the existence of a retarded 
Green's function is not guaranteed 
for an arbitrary ${\cal L}$; conditions 
(\ref{GGen}) single out equations 
that may have physical meaning. 

Although these definitions (as well as results below) 
are quite general, three cases are of major practical 
importance:
the nonrelativistic Schr\"odinger equation,  
\begin{eqnarray} 
\showlabel{SchEq}
{\cal L} = i\partial_t +  \frac{1}{2} \partial_r^2, 
\end{eqnarray}%
the wave equation,  
\begin{eqnarray} 
{\cal L} = \partial_t ^2 -  \partial_r^2, 
\end{eqnarray}%
and the relativistic Klein-Gordon equation,  
\begin{eqnarray} 
{\cal L} = \partial_t ^2 -  \partial_r^2 + 1, 
\end{eqnarray}%
where units were chosen so as to remove dimensional 
constants. 
Spatial Fourier transformation,  
\begin{eqnarray} 
 \psi (r,t) = \int dk\, \e{ikr} \psi _k(t), 
\end{eqnarray}%
reduces these equations to their single-mode cases: 
\begin{eqnarray} 
\showlabel{GSchr}
{\cal L} &=& i\partial_t -  \frac{k^2}{2} , \ \ 
G_k(t) = - i \theta (t) \e{-ik^2t/2}, 
\\
\showlabel{GWave}
{\cal L} &=& \partial_t ^2 + k^2, \ \ 
G_k(t) = \theta (t) \sin kt ,
\\
\showlabel{GKlG}
{\cal L} &=& \partial_t ^2  + k^2 + 1, \ \ 
G_k(t) = \theta (t)\sin t \sqrt{k^2 + 1} ,
\end{eqnarray}%
where the corresponding Green's functions 
are also given.

These Green's functions all have a singularity 
at zero time, so problems arise 
if the source $s$ is a singular function 
as well. 
E.g., if ${\cal L} = i\partial_t -  {k^2}/{2}$ and 
$s$ contains white noise, 
defining mathematically the inhomogeneous 
Schr\"odinger equations 
(\ref{BasicEq}) 
and (\ref{BasicInt}) requires 
specification of a  stochastic calculus \cite{Gardiner}. 
Another way around this problem is found by 
assuming that the Green's function and/or the 
source are {\em regularised\/}, e.g., properly 
smoothed, and a corresponding limiting procedure is employed  
for restoring their singular values. 
The choice of stochastic calculus then becomes a 
result of this regularisation. 
For example, regularising the source by introducing 
finite correlation times means that 
normal calculus holds at all stages of the regularisation 
procedure, so that in the end the Stratonovich calculus 
is recovered. 

As we shall argue now, 
regularising $G$ may lead to Ito calculus. 
Consider, for example, the equation,  
\begin{eqnarray} 
\showlabel{EqEx}
i \partial _t  \psi (t) =  \varepsilon \psi (t) W'(t) , 
\end{eqnarray}%
where
 $\varepsilon $ is a constant  and 
$W'$ is the derivative of the Wiener process. 
It can only be defined in the sense of the theory of 
generalised functions,  
\begin{eqnarray} 
\int dt W'(t) \varphi (t) = - \int dt W(t) \varphi '(t) , 
\end{eqnarray}%
where $\varphi (t)$ is a ``good'' function. 
It would suffice for $\varphi $ to have finite support 
and be continuously differentiable, so that $\varphi '$ is a 
function of finite variation. 
It is then clear that for $ \psi W'$ to be defined, 
it would suffice for $ \psi $ to be continuously 
differentiable, but this is certainly not consistent with  
Eq. (\ref{EqEx}). 

The standard way around this problem is to 
define $\int  \psi W' dt = \int  \psi dW$ as a  
stochastic integral so that the equation 
$ \psi  =  \psi (0) -i  \varepsilon 
\int _{0}^{t}  \psi  dW$ is 
defined. 
Instead, we replace an 
{\em undefined\/} integral equation, 
$ \psi =  \psi _0 - i  \varepsilon 
\int _{-\infty}^t  \psi W'dt'$, by a 
regularised integral equation,  
\begin{eqnarray} 
\showlabel{IntReg}
 \psi (t) =  \psi _0  +  \int G _{\text{reg}}(t-t')   
\psi (t') \varepsilon (t') W'(t') dt' .  
\end{eqnarray}%
$G _{\text{reg}}(t)$ is a regularised retarded 
Green's function, which is (i) causal, 
$G _{\text{reg}}(t) = 0, t \leq 0$, (ii) a given number 
of times continuously differentiable, and (iii) in a certain 
sense close to $G(t)$.
For example,    ($k \geq 1$)
\begin{eqnarray} 
\showlabel{GReg}
G _{\text{reg}}(t) = - i \theta (t)
\left ( 
1 - \e{- \Gamma  t}
\right )^{k + 1}
\end{eqnarray}%
is $k$ times continuously 
differentiable.
$\Gamma $ here is an  
arbitrary positive parameter, and the final limit 
$\Gamma \rightarrow \infty$ 
is implied. 
We have also introduced a truncating factor 
$\varepsilon (t)$ into the white noise factor $W'$:
$\varepsilon (t) \leq \varepsilon $ is infinitely differentiable, and 
a negative $T'$ exists such that 
$\varepsilon (t) = \varepsilon $ if $t > T'$ and 
$\varepsilon (t) = 0 $ if $t < 2 T'$; 
this is necessary to have a consistent $in$-field 
formulation. 
Eq. (\ref{IntReg}) is now  
consistent with the assumption of 
$ \psi $ being $k$ times continuously differentiable. 
If this is indeed the case, the factor 
$G _{\text{reg}}(t-t')   
\psi (t') \varepsilon (t')$ at a given $t > 2 T'$ is 
(at least)
continuously differentiable and has a finite 
support $\left \{ 
t':\ 2 T' < t' < t
\right \}$, so that the integral 
on the right of (\ref{IntReg}) is 
defined; then, continuity of its $k$-th derivative
 by $t$ follows from the 
fact that $W'$ is, loosely speaking, no more singular 
than a $ \delta $-function.  
(For $t < 2 T'$, (\ref{IntReg}) reduces to $ \psi  =  \psi _0$.)

An indication that in the limit 
$\Gamma \rightarrow \infty$ a solution to Eq. (\ref{IntReg})
approaches a solution to the Ito differential equation 
(\ref{EqEx}) may be seen from the following considerations. 
As a generalised function, $\varepsilon (t')W'(t')$ 
may be approximated by a discrete sum of $ \delta $-functions,  
\begin{eqnarray} 
\varepsilon (t')W'(t') \approx \sum _{k = 1}^{\infty}
\tilde \varepsilon _k  \delta (t - t_k) ,
\end{eqnarray}%
where $t_k = 2 T' + k  \Delta t$, and $  \Delta t$ 
is a discretisation scale. 
Eq. (\ref{IntReg}) then has a unique 
solution,  
\begin{eqnarray} 
\showlabel{ExSol}
 \psi (t) =  \psi _0 + \sum _{k = 1}^{\infty}
\tilde \varepsilon _k G _{\text{reg}}(t-t_k)  \psi (t_k) ,
\end{eqnarray}%
where $ \psi (t_k)$ may be found recurrently,  
\begin{eqnarray} 
 \psi (t_{m}) =  \psi _0 + \sum _{k = 1}^{m-1}
\tilde \varepsilon _k G _{\text{reg}}(t_{m}-t_k)  \psi (t_k) .
\end{eqnarray}%
The sum in (\ref{ExSol}) is in fact finite, so that 
$ \psi (t)$ inherits all the ``goodness'' of $G _{\text{reg}} (t)$. 
It is now easy to see that if 
$\Gamma  \Delta t \gg 1$, the integral in (\ref{IntReg}) 
coincides with the partial sum 
$-i \sum  \psi (t_k) \varepsilon (t_{k}) \left [ 
W(t_{k+1}) - 
W(t_{k}) 
\right ]$; 
so that in the limit $ \Delta t \rightarrow 0$ we recover 
the Ito integral 
$- i \int _{2 T'}^{t}dW(t' ) \psi (t')\varepsilon (t')$. 
From the practical point of view, this ``proves'' the hypothesis that in the limit 
$\Gamma \rightarrow \infty$ the Ito calculus is recovered,
since any real calculation 
implies time discretisation. 
In order to prove it mathematically, one should 
commute the  
limits:  
first $
\sum _{k = 1}^{\infty}
\tilde \varepsilon _k  \delta (t - t_k) 
\rightarrow 
\varepsilon (t')W'(t')  
$ and second 
$\Gamma  \rightarrow \infty $, 
while the above considerations imply the opposite 
order of the limits.

Below we shall see that $G(r,r',t-t')$ becomes 
the propagator in the diagram series so that regularising 
it 
prevents 
ultraviolet 
divergences 
in the diagrams 
(this is nothing but the Pauli-Villars regularisation
known in QFT \cite{PauliVillars}). 
The choice of stochastic 
calculus is hence connected to regularisations 
in the diagram series and hence to renormalisations. 
Investigation of this connection in full is a major 
undertaking; it should also include the continuous-space 
limit which we do not even try to tackle. 
At this stage, we just assume that  $G$ and/or 
$s$ are regular (or regularised if necessary), 
so that all our expressions 
make sense. 
Note that this means that the integral 
equation (\ref{BasicInt}) 
rather than the differential equation 
(\ref{BasicEq}) is considered:  the latter 
only emerges when regularisations are removed. 
\subsection{Characterisation of the stochastic sources and  fields} 
In order to have a fully defined stochastic differential equation,
the functional dependence of the source on the full 
({\em local\/} or {\em microscopic\/}) field, $ \psi (r,t)$, is
needed. Such a charcterisation of the source is usually given 
by the physical model considered. 
Then to  {\em solve\/} equations (\ref{BasicEq}) or 
(\ref{BasicInt}) means finding the field $\psi$ as function of the 
{\em in-\/}field $ \psi _0(r,t)$.

For example, in a homogeneous linear medium, 
$s(r,t) = \chi  \psi (r,t)$, 
where $\chi $ is the 
linear susceptibility of the medium.
In general,  when randomness is also 
included, the source is a stochastic variable
whose properties are determined by 
a probability distribution, $P(s| \psi )$, of the function 
$s$ {\em conditioned on the full field\/} $ \psi (r,t)$. 
It will be convenient to distingish between real and complex fields.
We will start from the former case and will generalize the results to
complex fields at the end.
We introduce a characteristic functional 
$S(\alpha | \psi )$ corresponding 
to $P(s| \psi )$ as,  
\begin{equation}
S(\alpha | \psi ) = 
\left . \overline{\text{e}^{\alpha s}} \right |_{ \psi } = 
\int Ds \, \e{\alpha s}P( s | \psi ), 
\end{equation}
Here, 
$\alpha (r,t)$ is an arbitrary ``good'' real function and 
 $\int Ds$ denotes a functional integration 
over the ``trajectories'' $s(r,t)$.  We use an 
abbreviated notation in which 
$\alpha s$ denotes $\int dx \alpha (x) s(x)$, where 
$x = \{r,t\}$ and $\int dx = \int dr dt$.
Differentiating $S$ over $\alpha (x)$ 
produces multi-space-time averages of the source, 
conditioned on the full field, 
\begin{equation}
\left . \overline{s(x_1)\cdots s(x_n)}\right  |_{ \psi } =
\frac{\partial ^n}{\partial \alpha (x_1)\cdots \partial \alpha (x_n)}
S(\alpha | \psi ) \Bigr|_{\alpha = 0}.
\end{equation}
To specify the functional dependence of the source on $\psi$ it is
convenient to introduce {\em generalised susceptibilities\/} 
(or just susceptibilities)
 $\chi ^{(m,n)}(x_1, \cdots , x_m;x'_1, \cdots , x'_n)$.
They are coefficients in the series expressing 
cumulants of the local source $s(x)$ in terms of the powers 
of the local field $ \psi (x)$. 
\begin{mathletters}
\showlabel{FullSource}
\begin{eqnarray} %
\nonumber  
\left . \overline{s(x)}\right |_{ \psi } &=& \chi ^{(1,0)}(x) 
+ \int dx' \chi ^{(1,1)}(x;x')  \psi (x') 
\\ && \hspace{17ex} 
+  \frac{1}{2} \int dx' dx'' \chi ^{(1,2)}(x;x',x'')  
\psi (x') \psi (x'')+\dots,
\\
\left . \overline{s(x)s(x')}\right  |_{ \psi } 
&=& 
\left . \overline{s(x)}\right |_{ \psi }
\left . \ \overline{s(x')}\right  |_{ \psi } 
+ \chi ^{(2,0)}(x,x') 
+ \int dx'' \chi ^{(2,1)}(x,x';x'')  \psi (x'')+\dots, 
\\ \nonumber & \vdots &
\end{eqnarray}
\end{mathletters}%
With them the generating functional can be given a simple form
\begin{eqnarray} 
S(\alpha | \psi ) = \exp \sum_{m,n=0}^{\infty} 
 \frac{1}{m!n!} \alpha ^m \chi ^{(m,n)}  \psi ^n ,
\end{eqnarray}%
where  
\begin{eqnarray} 
\alpha ^m \chi ^{(m,n)}  \psi ^n &=& 
\int 
dx_1\cdots dx_m
dx'_1\cdots dx'_n 
\times \nonumber  \\ &&
\alpha (x_1) \cdots \alpha (x_m)
\chi ^{(m,n)}(
x_1, \cdots , x_m
;
x'_1, \cdots , x'_n
)
 \psi (x'_1) \cdots  \psi (x'_n).  
\showlabel{ChiGen}
\end{eqnarray}%
The susceptibilities should obey the causality condition,   
\begin{eqnarray} 
\showlabel{CausChi}
\chi ^{(m,n)}(x_1, \cdots ,x_m;x'_1, \cdots ,x'_n) = 0, \ \ 
\max (t'_1, \cdots ,t'_n) > \max (t_1, \cdots ,t_m) ; 
\end{eqnarray}%
this condition can also be formulated as {\em the latest 
argument of a susceptibility is always an output one\/};
hence $\chi ^{(0,n)} = 0$ for all $n$. 

The quantity $\chi ^{(1,0)}(x)$ is a given source 
(non-random). 
For $m = 1$  
we find susceptibilities proper: the linear one, 
$ \psi ^{(1,1)}$,  and the nonlinear ones, 
$ \psi ^{(1,n)}$ for $ n > 1$. 
If $\chi ^{(m,n)} = 0$ for $ m > 1$, 
the dependence of the source on the local field is not 
stochastic, and equation (\ref{BasicEq}) is 
not stochastic either; it is linear, if $\chi ^{(1,n)} = 0$ for $n>1$, 
and otherwise nonlinear. 
Non-zero $\chi ^{(m,n)}$ for $ m>1$ 
introduce stochasticity: non-zero 
$\chi ^{(m,0)}$ are cumulants of  
a given  random source, 
while $\chi ^{(m,n)}$ for both $ m > 1$ and $ n> 0$ 
describe how the statistics of the source depends on 
the field. 

In practice, as a rule, only a finite number 
of generalised susceptibilities are non-zero. 
To be specific, we restrict ourselves to the case of 
only $\chi ^{(1,0)}$, $\chi ^{(1,1)}$, 
$\chi ^{(1,2)}$, $\chi ^{(2,0)}$ and $\chi ^{(2,1)}$ non-zero;  
these are exactly the susceptibilities shown explicitly in Eqs. 
(\ref{FullSource}). 
This case corresponds to a source which is Gaussian 
if conditioned on the full field  (this reservation is 
important).
Also, the susceptibilities are commonly 
local: e.g., $\chi ^{(1,1)}(x;x') \sim  \delta (x-x')$, 
so that the source at the point $x$ depends only 
on the local field at the same point. 
The causality condition (\ref{CausChi}) is then satisfied 
automatically. 
Locality gives more physical sense to the equations, 
but leads to mathematical problems. 
In this paper, we assume that 
 these problems    
are overcome 
by regularisations. 

Solving  equations (\ref{BasicEq}) and  (\ref{BasicInt}) 
means finding the field dependence on the $in$-field $\psi_0$.
It is immediately obvious that this is equivalent to finding
the dependence of the source on the in-field, i.e.~given $P(s\vert\psi)$
finding the probability distribution $\Pi(s\vert\psi_0)$.
Then, once the ``macroscopic'' probability distribution 
$\Pi( s | \psi _0)$  is 
known, solving for the field is  straightforward. 
To characterise the field, we introduce the  characteristic functional 
of the multi-space-time field averages,  
\begin{eqnarray} 
\Phi (\zeta ) = \Phi (\zeta |  \psi _0 )  = 
\left . \overline{\text{e}^{\zeta  \psi }} \right |_{ \psi _0} 
 ,
\end{eqnarray}%
so that 
\begin{eqnarray} 
\overline{ \psi (x_1)\cdots  \psi (x_n)} =  
\left . \overline{ \psi (x_1)\cdots  \psi (x_n)}\right |_{ \psi _0} =  
\frac{\partial ^n}{\partial \zeta (x_1)\cdots \partial \zeta (x_n)}
\Phi (\zeta ) |_{\zeta = 0}. 
\end{eqnarray}%
Since $ \psi  = Gs +  \psi _0$, the field statistics 
are effectively that of the source, and we have,  
\begin{eqnarray} 
\showlabel{PhiBySigma}
\Phi (\zeta ) =  
\int Ds \, \e{\zeta (G s +  \psi _0)}\Pi( s | \psi _0) =  
\e{\zeta \psi _0 }\Sigma( \zeta G | \psi _0) , 
\end{eqnarray}%
where 
$[\zeta G](x) = \int dx' \zeta (x') G(x', x)$.
Here we have introduced the characteristic functional $\Sigma(\alpha\vert
\psi_0)$ corresponding to $\Pi(s\vert\psi_0)$:
\begin{equation}
\Sigma (\alpha | \psi _0) = 
\left . \overline{\text{e}^{\alpha s}} \right |_{ \psi_0 } = 
\int Ds \, \e{\alpha s}\Pi( s | \psi _0) . 
\end{equation}

In the  complex case, 
our definitions must be generalised to allow one to consider 
averages containing the fields and sources as well as their complex
conjugates. 
Thus the notation $\Pi  (s| \psi _0)$ used for the real case, 
must 
be replaced 
for complex fields 
by 
$\Pi  (s,s^*| \psi _0,  \psi _0^*)$; the functional integration in the complex case 
will be denoted as $\int D \psi  D \psi ^*$ and $\int Ds Ds^*$, etc.  
The 
characteristic functionals $S$, $\Sigma $ and $\Phi $ 
are defined as, 
\begin{eqnarray} 
S(\alpha , \alpha \dg | \psi ,  \psi ^*) &=& 
\int Ds Ds^* 
\e{\alpha \dg s + \alpha s^*}P( s, s^* | \psi ,  \psi ^*), \\
\Sigma (\alpha , \alpha \dg | \psi _0,  \psi _0 ^*) &=& 
\int Ds Ds^* 
\e{\alpha \dg s + \alpha s^*}\Pi( s, s^* | \psi _0,  \psi _0 ^*) , \\
\Phi (\zeta,\zeta \dg ) = \Phi (\zeta,\zeta \dg |  \psi _0,  \psi ^* _0 )  
&=& 
\int Ds Ds^* 
\e{
\zeta \dg (Gs +  \psi _0 ) + 
\zeta (G^*s^* +  \psi _0 ^* )
}\Pi( s, s^* | \psi _0,  \psi _0 ^*) , 
\end{eqnarray}%
where $\alpha (x) , \alpha \dg (x), \zeta (x)$ and 
$\zeta \dg(x)$ are arbitrary ``good'' functions 
($\dg$ here is just a notation, not relevant to Hermitian 
conjugation). 
One can also consider $\alpha ,\alpha \dg$ and 
$\zeta , \zeta \dg$ as pairs of complex-conjugated 
functions, $\alpha \dg = \alpha ^*$, 
$\zeta \dg = \zeta ^*$. 
Then, e.g., 
\begin{eqnarray} 
\left . \overline{s^*(x_1)\cdots s^*(x_m)
s(x'_1)\cdots s(x'_n)}\right  |_{ \psi,  \psi ^* } &=& 
\frac{\partial ^{m+n}}
{\partial \alpha (x_1)\cdots \partial \alpha (x_m)
\partial \alpha \dg (x'_1)\cdots \partial \alpha \dg (x'_n)} 
\nonumber \\ &&  \times 
S(\alpha, \alpha \dg | \psi ,  \psi ^*) |_{\alpha = \alpha \dg = 0} . 
\end{eqnarray}%
The rest of the above relations  change accordingly. 
\subsection{Relation between $P(s| \psi )$ and $\Pi(s| \psi _0)$} 
Our goal now is to find a formal solution to Eq.\ (\ref{BasicEq}) in the form 
of a relation between the ``microscopic'' and ``macroscopic'' 
probability distributions, $P(s| \psi )$ and $\Pi(s| \psi _0)$. 
  Assume discretisation of the time axis, 
$s(t),  \psi (t) \rightarrow s(t_k),  \psi (t_k)$, where 
$t_k = k \Delta t, k= -\infty, \cdots, \infty$; 
a final limit of $ \Delta  t \rightarrow 0 $ is implied. 
We omit the spatial variable as irrelevant. 
With discretisation, averages of the source at the given 
field are given by a functional 
integration ($t,t', \cdots t''$ are among $t_k$),  
\begin{eqnarray} 
\left . \overline{s(t)s(t')\cdots s(t'')}\right |_{ \psi } = \int 
s(t)s(t')\cdots s(t'')P(s| \psi ) \prod_{k=-\infty}^{\infty}ds(t_k)
.
\end{eqnarray}%
The distribution $P(s| \psi )$ is causal, 
ie, $s(t_k)$ depends on $ \psi (t_m)$ only  for $m \leq k$  
(which is equivalent to the above causality condition for 
susceptibilities). 
This allows one to introduce  reduced probability 
distributions,  
\begin{eqnarray} 
P_m \left ( 
s _{\leq t_m}
\left  | \psi  _{\leq t_m}
\right ) \right . = \int 
P(s| \psi ) \prod_{k=m+1}^{\infty}ds(t_k), 
\end{eqnarray}%
where  
\begin{eqnarray} 
s _{\leq t_m} &=& \left \{ 
s(t_k):\ k\leq m
\right \}, \\
 \psi  _{\leq t_m} &=& \left \{ 
 \psi (t_k):\ k\leq m
\right \}.  
\end{eqnarray}%
The fact that $P_m $ depends on $\psi  _{\leq t_m}$ 
and not on the whole $ \psi $  leads to the following causality condition 
for the averages, 
\begin{eqnarray} 
\showlabel{PCausAv}
\frac{\partial }{\partial  \psi (t_m)} \,
\left . \overline{s(t)s(t')\cdots s(t'')}\right |_{ \psi } = 0, \ \ 
t_m > \max(t,t', \cdots t'').
\end{eqnarray}%

The unravelling of the source statistics in time 
is described by the conditional probability distribution,  
\begin{eqnarray} 
P_m \left ( 
s(t_m) \left  | 
s _{\leq t_{m-1}},
\psi  _{\leq t_m}
\right ) \right  .
= \frac{P_m \left ( 
s _{\leq t_m}
\left  | \psi  _{\leq t_m}
\right ) \right  .
}{P_{m-1} \left ( 
s _{\leq t_{m-1}}
\left  | \psi  _{\leq t_{m-1}}
\right ) \right  .
} 
\end{eqnarray}%
In turn,  
\begin{eqnarray} 
P_m \left ( 
s _{\leq t_m}
\left  | \psi  _{\leq t_m}
\right ) \right  . = \prod_{k = -\infty}^{m}
P_k \left ( 
s(t_k) \left  | 
s _{\leq t_{k-1}},
\psi  _{\leq t_k}
\right ) \right  .
\end{eqnarray}%
Since we always observe a system only 
for finite times, $ P_m $ with $m$ large 
enough is actually as good as $P(s| \psi )$, 
so that we can write  
\begin{eqnarray} 
\showlabel{Unravel}
P(s| \psi ) = \prod_{k = -\infty}^{\infty}
P_k \left ( 
s(t_k) \left  | 
s _{\leq t_{k-1}},
\psi  _{\leq t_k}
\right ) \right  .
\end{eqnarray}%
(More rigorously, this relation implies $w$-limits 
for both plus and minus infinity.)

The advantage of the ``unravelling'' 
representation (\ref{Unravel}) for $P(s| \psi )$ 
is that it is perfectly designed so as to accept the dynamical 
relation between the source and the field, 
Eq.\ (\ref{BasicInt}). 
With time discretisation, it is understood as, 
\begin{eqnarray} 
\showlabel{GDiscr}
 \psi (t_m) =   \psi _0(t_m) + 
\Delta t \sum _{k = -\infty}^{\infty}G(t_m - t_k) s(t_k). 
\end{eqnarray}%
Note that since $G(t)$ is both causal and 
regular (or regularised), $G(0) = 0$ and hence 
the latest source value to contribute 
to $\psi (t_m)$ is $s(t_{m-1})$, 
ie, $\psi (t_m)$ depends on $s_{\leq t_{m-1}}$ 
($ \psi _0$ is either non-random 
or uncorrelated with $s$). 
Then, both $s_{\leq t_{m-1}}$ and 
$ \psi _{\leq t_{m}}$ in the conditional 
probability $P_m \left ( 
s(t_m) \left  | 
s _{\leq t_{m-1}},
\psi  _{\leq t_m}
\right ) \right  .
$ are effectively non-random and 
we can write  
\begin{eqnarray} 
\showlabel{SelfActCond}
\Pi_m \left ( 
s(t_m) \left  | 
s _{\leq t_{m-1}}, \left ( 
\psi _0
\right )_{\leq t_m}
\right ) \right  . 
= 
P_m \left ( 
s(t_m) \left  | 
s _{\leq t_{m-1}},
\left ( 
Gs+ \psi _0
\right )  _{\leq t_m}
\right ) \right  . .
\end{eqnarray}%
Here, $\Pi_m \left ( 
s(t_m) \left  | 
s _{\leq t_{m-1}}, \left ( 
\psi _0
\right )_{\leq t_m}
\right ) \right  .$ 
is the probability distribution 
for the source at $t_m$, conditioned 
on its own prehistory and the $in$-field. 
Relation (\ref{SelfActCond}) solves 
the self-action problem for the source, 
expressing its actual 
(macroscopically observable) statistics 
in terms of the (microscopic) relations 
characterising the system. 
For the multi-time probability 
distribution for the source, conditional 
on the $in$-field,  we have, 
\begin{eqnarray} 
\Pi  (s| \psi _0) = \prod_{m=-\infty}^{\infty}
\Pi_m \left ( 
s(t_m) \left  | 
s _{\leq t_{m-1}}, \left ( 
\psi _0
\right )_{\leq t_m}
\right ) \right  . ,
\end{eqnarray}%
and hence,  
\begin{eqnarray} 
\showlabel{SelfActFull}
\Pi  (s| \psi _0) = P( s | Gs +  \psi _0). 
\end{eqnarray}%
This relation originates in Eq.\ (\ref{BasicEq}) 
(regularised!) and causality.

The above derivation does not depend on whether we deal with a 
real or complex field. 
However, following the above convention 
regarding notation, 
 in the complex case it 
will be written as  
\begin{eqnarray} 
\showlabel{SelfActFullCmpl}
\Pi  (s,s^*| \psi _0, \psi _0^*) = 
P( s, s^* | Gs +  \psi _0, G^*s^* +  \psi _0^*).   
\end{eqnarray}%
Relation (\ref{SelfActFull}) applies in the real case. 

\subsection{Causal Wick's theorems} 
Our aim is now to rewrite relation (\ref{SelfActFull}) in 
terms of the characteristic functionals. To this end, 
we first rewrite this relation using the shift operator,  
\begin{eqnarray} 
\Pi(s| \psi _0) = \exp\left ( 
\frac{\partial }{\partial  \psi }Gs
\right ) P(s| \psi ) |_{ \psi = \psi _0}. 
\end{eqnarray}%
Here, we again use condensed  notation, 
$\frac{\partial }{\partial  \psi }Gs = 
\int dx dx' \frac{\partial }{\partial  \psi (x)}G(x;x')s(x')$;
note that the spatial dependence of the 
fields is restored.
Then,  
\begin{eqnarray} %
\nonumber  %
\Sigma (\alpha  | \psi _0) &=& 
\int Ds \exp\left ( \alpha s + 
\frac{\partial }{\partial  \psi }Gs
\right ) P(s| \psi ) |_{ \psi = \psi _0} 
\\ &=&
\exp\left (
\frac{\partial }{\partial  \psi }
G
\frac{\partial }{\partial \alpha }
\right )
\int Ds \, 
\e{\alpha s} P(s| \psi ) |_{ \psi = \psi _0} 
=
\exp\left (
\frac{\partial }{\partial  \psi }
G
\frac{\partial }{\partial \alpha }
\right ) S(\alpha | \psi ) |_{ \psi  =  \psi _0} .
\showlabel{WickSourceReal}
\end{eqnarray}%

It is worth noting why we have to write, 
$\frac{\partial }{\partial  \psi }(\cdots)|_{ \psi = \psi _0}$, rather 
than just 
$\frac{\partial }{\partial  \psi _0}
$. 
The problem is that $ \psi _0$ is a solution 
to a free equation, 
ie, $\frac{\partial }{\partial   \psi _0}$ is  
a derivative with constraints, 
whereas when applying 
the shift operator, 
$\exp\left ( 
\frac{\partial }{\partial  \psi }Gs
\right )
$, to $P(s| \psi )$ in order 
to turn it into $P(s| Gs + \psi )$, 
one has to assume that $ \psi $ is arbitrary. 
Since $P(s| \psi )$ 
is indeed defined for an arbitrtary $ \psi $, 
the whole situation is consistent.

Relation (\ref{WickSourceReal}) is startlingly 
reminiscent of relations known 
in QFT that 
express Wick's theorem for bosonic operators 
as a differential operation \cite{Akhiezer,Hori}, 
and we shall call it the 
{\em causal Wick's theorem\/}. 
Below we shall see that it plays the same role in 
deriving diagram series for the SDEs as Wick's 
theorem proper plays in deriving diagram series 
for interacting bosonic fields. 
The assumptions it is based on are: (i) causality, 
(ii) Eq. (\ref{BasicInt}) and (iii) the condition 
$G(r,r',0) = 0$. 
Note that the last condition is meaningful only 
for regularised SDEs. 

For the functional $\Phi(\zeta )$, 
relations (\ref{WickSourceReal}) and (\ref{PhiBySigma}) 
result in, 
\begin{eqnarray} 
\showlabel{WickFieldReal}
\Phi (\zeta ) = \exp\left ( 
\frac{\partial }{\partial  \psi }
G
\frac{\partial }{\partial \alpha }
\right )
\exp\left ( 
\zeta  \psi 
\right )
S (\alpha | \psi )|_{\alpha =  0, \psi =  \psi _ 0}.
\end{eqnarray}%
This relation will be referred to as a 
{\em generating formula for causal diagram series\/}. 

In the complex case, relations 
expressing $\Sigma$ and $\Phi $ over $S$ 
(i.e., the causal Wick's theorem and the generating formula for causal 
series)
are found to be, 
\begin{eqnarray} 
\showlabel{WickSourceCmpl}
\Sigma (\alpha, \alpha \dg  | \psi _0,  \psi _0 ^*) &=& 
\exp\left (
\frac{\partial }{\partial  \psi }
G
\frac{\partial }{\partial \alpha \dg}
+
\frac{\partial }{\partial  \psi \dg}
G^*
\frac{\partial }{\partial \alpha }
\right ) S(\alpha, \alpha \dg  | \psi ,  \psi  \dg ) 
|_{ \psi  =  \psi _0, \psi  \dg =  \psi _0 ^*} , 
\\
\showlabel{WickFieldCmpl}
\Phi (\zeta ,\zeta \dg) &=& \exp\left ( 
\frac{\partial }{\partial  \psi }
G
\frac{\partial }{\partial \alpha \dg}
+
\frac{\partial }{\partial  \psi \dg }
G^*
\frac{\partial }{\partial \alpha }
\right )
\exp\left ( 
\zeta  \dg \psi 
+
\zeta  \psi \dg
\right )
 \times \nonumber \\ && \ \ \ \ 
S (\alpha , \alpha \dg| \psi ,  \psi \dg )
|_{\alpha =  \alpha \dg =  0, \psi  =  \psi _0, \psi  \dg =  \psi _0 ^*}.
\end{eqnarray}%
In these relations, $ \psi ,  \psi \dg$ is a pair of 
``good'' functions which can be either arbitrary or complex 
conjugated. 
We have assumed that $S (\alpha , \alpha \dg| \psi ,  \psi ^* )
$ may be regarded as an analytic function of separately 
$ \psi $ and $ \psi ^*$; in all practical examples $S = \exp 
(\text{Polynomial of } \psi , \psi ^*)$ 
(cf Eq.\ \ref{FullSource}), so that this assumption 
is valid. 
Note, without going in detail, that assuming  
$ \psi , \psi \dg$ ``good'' is consistent only with 
a  regular $G(x;x')$, so that relations (\ref{WickSourceCmpl}) 
and (\ref{WickFieldCmpl}) imply regularisations; 
this certainly applies to  (\ref{WickSourceReal}) 
and (\ref{WickFieldReal}) as well. 

\section{Diagram expansions and SDE\mbox{s}}
In QFT, relations similar to 
the causal Wick's theorems are used in order to express 
Wick's theorem proper 
in a compact form as a differential operation \cite{Akhiezer,Hori}. 
Then, if this form of Wick's theorem is plugged into the standard 
perturbation approach \cite{Hori,II}, 
relations for Green's functions of the interacting bosonic 
fields are found which are structurally identical to relations  
(\ref{WickFieldReal}) or (\ref{WickFieldCmpl}) for classical averages 
(this identity is straightforward for non-relativistic quantum fields
\cite{OPO,Feedback}, but becomes more involved in 
relativistic QFT \cite{Corresp}). 
In principle, this identity is all that is necessary 
in order to establish the link 
between the quantum and classical field theories. 
For this purpose, diagram series (which both in QFT and CFT  may be found 
by expanding generating formulae in power series) are redundant. 
Moreover, establishing the quantum/classical link in terms 
of the diagram series, 
strictly speaking, degrades the approach since it brings over the  
problem of convergence of the diagram series and confines  
the link to the perturbation theory (whereas, e.g., for the 
anharmonic oscillator linking the generating formulae 
leads to {\em exact\/} results \cite{PRL,II}). 

Despite all this, we believe introducing diagram series in CFT 
may be beneficial. First, diagrams are common language in QFT, and 
introducing them in CFT eliminates the seeming incompatibility 
between  the q- and c-number techniques. 
Second,  
diagram expansions in CFT are not introduced as a computational tool. 
They are regarded only as a leading consideration, 
visualising certain structural properties of an SDE and/or 
the respective generating relation. These properties should then be proved directly. 
(Similarly, in QFT analysing diagram 
series leads to relations 
for observable quantities, like Dyson equations. 
As a rule, these relations may be derived independently.)
Hence  
rigor of the diagram approach is not an issue. 
Third, in view of the encountered link between regularisations 
and stochastic calculus, introducing diagrams related to SDEs 
becomes a must: 
after all, those are diagrams that diverge!
\subsection{Causal diagram series}
For simplicity, we confine ourselves to a real SDE, 
with the noise 
source specified 
by the set of susceptibilities shown explicitly 
in Eqs. (\ref{FullSource}). Examples of series related to 
complex SDEs 
will be presented elsewhere \cite{II}.

Graphically,  the quantities $ \psi _0(x)$ and 
$G(x;x')$ will be denoted as lines,  
\begin{mathletters}
\showlabel{BasicLines}
\begin{eqnarray} 
\showlabel{GReal}
G(x;x') &=& \vframe{8}{3.5}{0}{0}{%
\lnd{0}{0}{8}{$\! \! x'$}{$\ \ \ x$}{}
\tarr{2}{-2}{4}
}
, \\ 
\showlabel{rPsi0}
\psi  _0(x) &=& \vframe{4.5}{3.5}{0}{0}{%
\lnd{0}{0}{4}{}{$\ \ \ x$}{}
\tarr{0.25}{-2}{4}
}  , 
\end{eqnarray}%
\end{mathletters}%
while $\zeta (x)$ and the susceptibilities 
$\chi ^{(m,n)}$ as vertices,  
\begin{mathletters}
\showlabel{BasicVertices}
\begin{eqnarray} 
\zeta (x) &=& 
\dframe{4}{3}{\subp{0}{0}{%
\txt{0}{1}{$x$}
\dotted\linePT{0}{0}{4}{0}{\vrt{0}{0}}
}
\tarr{0}{-2}{4}
}
,  
\showlabel{ZetaReal}
\\
\chi ^{(1,0)} (x) &=& 
\dframe{4}{3}{\subp{0}{0}{%
\vrt{0}{0}
\dotted\linePT{0}{0}{4}{0}{\txt{-0.5}{1}{$x$}}
}
\tarr{0}{-2}{4}
}
,
\showlabel{s0Real}
\\
\showlabel{Chi11}
\chi ^{(1,1)}(x;x') &=& \figPTaj ,
\\
\chi ^{(1,2)}(x;x',x'') &=& 
\dframe{8.5}{4}{\subp{4.5}{0}{%
\vrt{0}{0}
\dotted
\linePT{0}{0}{4}{0}{\txt{-0.5}{1}{$x$}}
\linePT{0}{0}{-\dd}{\dd}{\txt{-2}{-1}{$x'$}}
\linePT{0}{0}{-\dd}{-\dd}{\txt{-2}{0}{$x''$}}
}
\tarr{3.5}{-2.5}{4}
},
\\ 
\chi ^{(2,0)}(x,x') &=& \figPTbf , 
\\
\chi ^{(2,1)}(x,x';x'') &=& \figPTbj .
\showlabel{Chi21}
\end{eqnarray}%
\end{mathletters}%
The ``time arrow'', which is drawn below 
each diagram, distinguishes graphically input and 
output arguments of the lines and vertices:   
The argument of $ \psi _0(x)$
is  regarded 
as a {\em line output\/}, as well as  
the  ``future'' argument (i.e., $x$) 
in $G(x;x')$; 
the ``past'' 
argument (i.e., $x'$) in $G(x;x')$ is regarded 
as a {\em line input\/}. 
A generalised susceptibility, $\chi ^{(m,n)}(
x_1, \cdots , x_m
;
x'_1, \cdots , x'_n
)$, is a quantity with $n$ 
{\em vertex inputs\/} and 
$m$ {\em vertex outputs\/}. 
By definition, the argument of  $\zeta (x)$ 
is a vertex input. 

We now expand all exponents in the generating relation 
(\ref{WickFieldReal}) in power series, and consider a certain 
term in these series. (In detail, this procedure is discussed in the appendix.) 
We see that $\alpha (x)$ always occurs convolved with a vertex output, 
while $\frac{\partial }{\partial \alpha (x)}$ is always convolved with a 
line input. Since $\frac{\partial \alpha (x)}{\partial \alpha (x')} =  \delta (x-x')$, 
differentiating by $\alpha $'s leaves all  vertex outputs pairwise 
convolved with line inputs. 
No free vertex outputs or line inputs may remain; 
terms with unequal number of these give zero. 
Similarly, derivatives $\frac{\partial }{\partial  \psi (x)}$ leave 
vertex inputs convolved with the line outputs of the 
propagators $G(x;x')$; ``surviving'' $ \psi $'s become $ \psi _0$'s. 
As a result, we find  all  vertex inputs pairwise 
convolved with line outputs, and no free arguments remain. 
Graphically, convolved input-output pairs are denoted by 
connecting respective ends of the lines to the vertices, e.g.,  
\begin{mathletters}
\showlabel{CompleteCaus}
\begin{eqnarray} 
\showlabel{ZetaPsi0}
\vframe{4.5}{3}{0}{0}{%
\lnd{0}{0}{4}{}{}{}
\vrt{4}{0}
\tarr{0.25}{-2}{4}
} &=&  
\int dx \zeta (x)  \psi _0 (x), 
\\
\showlabel{ZetaGs0}
\vframe{8}{3}{0}{0}{%
\lndrs{0}{0}{8}{}{}{}
\tarr{2}{-2}{4}
}
 &=&  
\int dx dx' \zeta (x) G(x;x')  \chi ^{(1,0)}(x') 
,  
\\
\figPTbl &=& \int dx_1 \cdots dx_5
\zeta (x_1) \zeta (x_2) 
G(x_1;x_3) G(x_2;x_4) 
\nonumber  \times \\ &&
\chi ^{(2,1)}(x_3,x_4;x_5)
 \psi _0(x_5) , 
\\
\figPTbp &=& \int dx_1 \cdots dx_6
\zeta (x_1) G(x_1;x_2) 
\chi ^{(1,2)}(x_2;x_3,x_4)
\nonumber  \times \\ &&
G(x_3;x_5) G(x_4;x_6) 
\chi ^{(2,0)}(x_5,x_6) .
\end{eqnarray}%
\end{mathletters}%
Expressions of such structure will be called  
{\em causal diagrams\/}. 
Note that the ``time 
arrow'' applies to all elements in a  diagram. 

In general, a causal diagram is a product of the basic elements---%
lines and vertices, where some line inputs and outputs are pairwise convolved 
with, respectively,  vertex outputs and inputs. 
A diagram containing free arguments is called {\em incomplete\/}; 
e.g., (\ref{BasicLines}) and (\ref{BasicVertices}) are legitimate incomplete 
diagrams. 
A diagram without free arguments (cf (\ref{CompleteCaus})) is {\em complete\/}. 
As is shown in the appendix, the functional $\Phi(\zeta )$ 
is expressed as a sum of all complete causal diagrams which may be built of 
the elements (\ref{GReal}--\ref{Chi21}), with certain 
coefficients; the rules for calculating these coefficients may be found in textbooks. 
Incomplete diagrams occur, e.g., in diagram expansions 
for the field averages.  Formally, these are found by ``stripping'' 
the $\zeta $ vertices from the complete diagrams.  
The simplest example of such a diagram is (\ref{rPsi0}): 
it  contributes  
to $\overline{ \psi (r,t)}$ and  
is found by stripping the $\zeta $ vertex from (\ref{ZetaPsi0}). 

It is also worth 
noting that whereas the ``time flow'' in diagrams is 
from left to right, in analytical expresssions, 
as a rule, time increases from right to left.
Eg, in the diagram in (\ref{ZetaGs0}) the 
$\zeta $ vertex is on the right,
whereas in the analytical expression 
$\zeta G \chi ^{(1,0)} = 
\int dx dx' \zeta (x) G(x;x')  \chi ^{(1,0)}(x') $ 
the natural position 
of the function  $\zeta (x) $ is on the left. 
The diagram notation thus commonly has an inverted 
order of objects compared with the analytical 
notation. 
\subsection{Connected diagrams and field cumulants} \label{cums}
A diagram that graphically consists of a 
number of separate sub-diagrams without 
common elements is called {\em disconnected\/};
otherwise, a diagram is {\em connected\/}. 
A disconnected diagram is a product of its 
connected components. 
In the above examples, all diagrams are connected;  
however, full causal diagram series for the functional 
$\Phi(\zeta )$ contain {\em all\/} possible complete diagrams, 
connected as well as disconnected.

To get rid of disconnected diagrams, one should describe 
the field in terms of its {\em  cumulants\/} rather than averages. 
Formally, the field cumulants $C^{(n)}, n = 0, 1, \cdots$, 
are defined as, 
\begin{eqnarray} %
\showlabel{RealByCum}
\Phi (\zeta ) = 
\exp \sum _{n = 1}^{\infty} 
\frac{1}{n!} \zeta ^n C^{(n)} ,
\end{eqnarray}%
where ($n = 1, 2, \cdots$)
\begin{eqnarray} 
\zeta ^n C^{(n)} &=& 
\int dx_1 dx_2 \cdots dx_n 
C^{(n)}(x_1,x_2,\cdots , x_n)
\zeta (x_1) 
\zeta (x_2) \cdots 
\zeta (x_n) . 
\end{eqnarray}%
In particular, 
\begin{eqnarray} %
 \overline{ \psi (x) } &=& C^{(1)}(x) , \\
\overline{ \psi (x)  \psi (x') } &=& C^{(2)}(x,x') 
+ C^{(1)}(x)C^{(1)}(x') , \\
\overline{ \psi (x)  \psi (x')   \psi (x'') } &=& 
C^{(3)}(x,x',x'')  
+ C^{(2)}(x,x') C^{(1)}(x'')
+ C^{(2)}(x,x'') C^{(1)}(x')
\nonumber \\ &&
+ C^{(2)}(x',x'') C^{(1)}(x) 
+ C^{(1)}(x)C^{(1)}(x')C^{(1)}(x''), 
\\ \nonumber &\vdots& 
\end{eqnarray}%
i.e., expression of a particular field average 
by the cumulants corresponds to its 
all possible factorisations. 
Characterisation of the field statistics by 
cumulants is more economical than characterisation 
by field averages.
E.g., if $ \psi (x)$ is non-random, $C^{(1)}(x) =  \psi (x)$, 
and $C^{(n)} = 0, n>1$; 
if $ \psi (x)$ is random and Gaussian, 
its cumulants vanish for $n>2$; 
whereas all field averages are non-zero even 
for non-random $ \psi $.

There exists a general theorem (Mayer's first theorem) 
stating that the diagram expansion of the logarithm 
$\Phi(\zeta )$ contains only connected diagrams,  
so that 
\begin{eqnarray} 
\Phi(\zeta ) = \exp [\text{conn}\Phi (\zeta )] ,
\end{eqnarray}%
where $\text{conn} \Phi (\zeta )$ is given by diagram series 
where all disconnected diagrams are dropped while connected 
ones retain their coefficients. 
Comparing this with the definition of the field cumulants, 
we see that  
\begin{eqnarray} 
\text{conn}\Phi (\zeta ) = \ln \Phi (\zeta ) = \sum _{n = 1}^{\infty} 
\frac{1}{n!} \zeta ^n C^{(n)} .
\end{eqnarray}%
Hence (cf the examples below) a diagram expansion for 
$\frac{1}{n!} \zeta ^n C^{(n)}$ 
contains all connected complete diagrams with exactly $n$ 
$\zeta $-vertices (\ref{ZetaReal}), occurring with the coefficients they had in the series 
for $\Phi(\zeta )$. 
Correspondingly, the expansion for $C^{(n)}\left ( 
x_1, \cdots , x_n
\right )$ contains all  connected incomplete causal diagrams 
with exactly $n$ free line outputs. 
\subsection{Diagram structures corresponding to 
certain types of equations} 
\subsubsection{Emission of  given sources} 
To start with, we consider the simplest possible 
stochastic problem of radiation of a {\em given random\/} 
source, $s(x) = s_0(x)$. 
For simplicity (and also to stay within (\ref{FullSource})), 
we assume that it is Gaussian, and described 
by the cumulants $\overline{s_0(x)} 
= \chi ^{(1,0)}(x)$ and $
\overline{s_0(x)s_0(x')} - 
\overline{s_0(x)}\ \overline{s_0(x')}
= \chi ^{(2,0)}(x,x')$   
(conditioning on the full field 
is irrelevant for a given source). 
This problem is readily solved,  
\begin{eqnarray} 
\psi (x) 
&=&  \psi _0 (x) +  \psi '(x) , \\
\psi '(x) &=& \int dx' G(x;x') s_0(x'), 
\end{eqnarray}%
where $\psi '(x)$ is the emitted field. 
For the field cumulants we have,  
\begin{mathletters}
\showlabel{CumGiven}
\begin{eqnarray} 
C^{(1)}(x) &=&
 \psi _0(x) + \int dx' G(x;x') \chi ^{(1,0)}(x') 
=  \figPTap +  \figPTas
, \\ 
C^{(2)}(x,x') &=& 
\int dx'' dx''' G(x;x'') G(x';x''') 
\chi ^{(2,0)}(x'',x''') 
=  \figPTbi 
,
\end{eqnarray}%
\end{mathletters}%
so that, 
\begin{eqnarray} 
\showlabel{PhiGiven}
\ln \Phi (\zeta ) =  
\zeta C^{(1)} +
\frac{1}{2} \zeta ^2 C^{(2)} = 
\figPTac + \figPTai +  \frac{1}{2} \figPTbh . 
\end{eqnarray}%

It is easy to see that 
these are indeed the three 
(and only three)
connected complete diagrams 
that can be built of 
the available graphical elements 
$ \psi _0,\ G,\ \chi ^{(1,0)}$ and $ \chi ^{(2,0)}$, 
with right coefficients (cf the appendix).  
Hence expressions (\ref{CumGiven}) and (\ref{PhiGiven}) 
are exactly that one would 
find from the diagrammatic approach. 
Note, however, that the logic of the diagrammatic solution 
is reversed: 
relation  (\ref{PhiGiven}) is found summing 
all legitimate complete connected diagrams, 
while relations (\ref{CumGiven}) follow. 
\subsubsection{Linear susceptibility}
Consider now an equation with a non-zero 
linear susceptibility, 
$\chi ^{(1,1)}(x;x') \neq 0$. 
For simplicity, we assume that the only other non-zero 
susceptibility is $\chi ^{(1,0)}(x)$, 
i.e., we consider radiation of a given non-random source 
into a linear medium. 
The following considerations are nevertheless applicable 
in a general case of any set of non-zero susceptibilities.

It is easy to see that all connected diagrams 
containing only the vertices $\chi ^{(1,1)}(x;x')$ 
and $\chi ^{(1,0)}(x)$ are 
{\em chains\/},  
\begin{eqnarray} 
\showlabel{RealChains}
\ln \Phi(\zeta ) &=& \figPTac + \figPTal + \figPTan + \cdots + 
\nonumber \\ && 
\figPTai + \figPTam + \figPTao + \cdots . 
\end{eqnarray}%
On their  ``past'' end,  
the chains are terminated either by the line $ \psi _0$, 
\begin{eqnarray} %
\figPTac &=& \zeta  \psi _0 , 
\\
\figPTal &=& \zeta  G \chi ^{(1,1)}\psi _0 , 
\\
\figPTan &=& \zeta  G \chi ^{(1,1)}G \chi ^{(1,1)} \psi _0 , 
\\
&\vdots&
\end{eqnarray}
or by the vertex $s_0$,   
\begin{eqnarray}
\figPTai &=& \zeta  G \chi ^{(1,0)} , 
\\
\figPTam &=& \zeta  G \chi ^{(1,1)} G \chi ^{(1,0)} , 
\\
\figPTao &=& \zeta  G \chi ^{(1,1)} G \chi ^{(1,1)} G \chi ^{(1,0)} , 
\\
&\vdots &
\end{eqnarray}%
The coefficients of the chains are all 
equal to one. 
(Formally, there is an additional 
class of 
connected diagrams, namely,  loops, 
\begin{eqnarray} %
\nonumber  
&&
\dframe{4}{3.7}{\subp{2}{0.5}{%
\thicklines 
\put(0,0){\oval(4,4)}\vrt{0}{-2}
}
\tarr{0}{-3}{4}
} + 
 \frac{1}{2} 
\dframe{8}{3.7}{\subp{4}{0.5}{%
\thicklines 
\put(0,0){\oval(8,4)}
\vrt{-2}{-2}
\vrt{2}{-2}
}
\tarr{2}{-3}{4}
} + 
 \frac{1}{3} 
\dframe{12}{3.7}{\subp{6}{0.5}{%
\thicklines 
\put(0,0){\oval(12,4)}
\vrt{0}{-2}
\vrt{-4}{-2}
\vrt{4}{-2}
}
\tarr{4}{-3}{4}
} + \cdots = 
\\
&&
\text{Tr} G\chi ^{(1,1)} + 
 \frac{1}{2} \text{Tr} G\chi ^{(1,1)}G\chi ^{(1,1)} + 
 \frac{1}{3} \text{Tr} G\chi ^{(1,1)} G\chi ^{(1,1)} G\chi ^{(1,1)} + 
\cdots .  
\end{eqnarray}%
They are zero due to causality conditions 
and regularisations. )

There are two possibilities as to how $\chi ^{(1,1)}$ 
can appear in a diagram: (i) between two $G$s, and (ii) 
between a $ \psi _0$ and $G$; 
it is the former that is responsible for infinite number of 
chains.  
Then, consider the sum of chains,  
\begin{eqnarray} 
\showlabel{GChains}
G'(x;x') &=& \figPTba +  \figPTbb +  \figPTbc + \cdots 
\\
&=& G(x;x') + \int dx''dx'''G(x;x'')
\chi ^{(1,1)}(x'';x''')G(x''';x') + \cdots . 
\end{eqnarray}%
It obeys an integral (Dyson) equation,  
\begin{eqnarray} 
G' = G + G\chi ^{(1,1)}G' .
\end{eqnarray}%
Acting on it by the operator ${\cal L}$, and using that 
${\cal L}G = \openone$, we find,  
\begin{eqnarray} 
{\cal L}' G' = \openone , 
\end{eqnarray}%
where 
${\cal L}' = {\cal L} - \chi ^{(1,1)}$. 
Thus partial summation of the chains in diagrams 
corresponds to shifting the linear susceptibility from the 
source to the ``free'' equation.

Replacing $G \rightarrow G'$ allows one to drop 
all diagrams containing a $\chi ^{(1,1)}$ vertex between 
two $G$ lines. 
As to the $\chi ^{(1,1)}$ vertices placed between $ \psi _0$s and 
$G$s, 
these are only found in the combination,  
\begin{eqnarray} 
\showlabel{Psi0Mod}
\psi _0 '(x) = \figPTap + \figPTaq , 
\end{eqnarray}%
where the propagator is now $G'$.  
It is easy to see that ${\cal L} \psi _0 = 0$ 
results in ${\cal L}' \psi _0 ' = 0$. 
Hence by redefining the graphical notation,  
\begin{eqnarray} 
 \figPTap &=& \psi _0 '(x) , \\
 \figPTt &=& G'(x;x') , \\
 \figPTaj &=& 0 , 
\end{eqnarray}%
(and dropping the primes) we arrive at an equivalent 
problem where 
the linear susceptibility is included into the operator 
${\cal L}$. 
Below we always assume this to be the case.

Note that instead of redefining the $in$-field, 
one could equally redefine the $\chi ^{(1,0)}$ 
vertex,  
\begin{eqnarray} 
\showlabel{s0Mod}
\figPTbe + \figPTbd  
\rightarrow  
\figPTbe , 
\end{eqnarray}%
leaving $ \psi _0$ unchanged. 
Then $ \psi _0$ is no longer  
a solution to the ``free'' equation, but, 
firstly, 
when deriving diagrams using Eq.\ (\ref{WickFieldReal})  
this fact is irrelevant, 
secondly, ${\cal L} \psi _0 = 0$, and hence 
${\cal L}' \psi _0 ' = 0$, may anyway 
be not the case after 
regularisations. 
\subsubsection{Nonlinear non-stochastic equations}
We now consider an equation
with quadratic nonlinearity, 
$\chi ^{(1,2)} \neq 0$. 
We assume that this equation is  non-stochastic, 
$\chi ^{(2,0)} = 
\chi ^{(2,1)} = 0$, and that $\chi ^{(1,1)} = 0$
(or 
included in the operator ${\cal L}$). 
The available graphical elements hence are  
$ \psi _0, G, \chi ^{(1,0)}$ and $\chi ^{(1,2)}$;
it is then easy to see that, with two exceptions, 
 all connected 
diagrams that one can build 
using these elements are trees branching into the past:
\begin{eqnarray} 
\ln \Phi(\zeta ) &=&
 \figPTac + 
 \figPTai  
\nonumber \\ && 
+ \frac{1}{2} \figPTav
+ \figPTax
+ \frac{1}{2} \figPTaz
\nonumber \\ && 
+ \frac{1}{2} \figPTay 
+ \frac{1}{8} \figPTaw + \cdots .
\showlabel{RealTrees}
\end{eqnarray}%
The futuremost vertex in the trees is $\zeta $, 
branching occurs at $\chi ^{(1,2)}$ vertices, and 
branches are ultimately terminated by 
$ \psi _0$ lines or 
$\chi ^{(1,0)}$ vertices. 
The series (\ref{RealTrees}) contain 
all such trees, each with a coefficient 
$1/g$, where $g$ is the order of the group of symmetry 
of the tree. 
\subsubsection{Stochastic differential equations}
Consider firstly a special case when $\chi ^{(m,n)} = 0$ 
for $n > 1$. 
Connected diagrams are then trees branching into the future:  
\begin{eqnarray} 
\showlabel{TreesRealStoch}
\ln \Phi (\zeta ) &=& 
\figPTac + \figPTai 
+  \frac{1}{2} \figPTbh 
+  \frac{1}{2} \figPTbl 
\nonumber  \\ && 
+  \frac{1}{2} \figPTbm 
+  \frac{1}{2} \figPTbn 
+  \frac{1}{8} \figPTbo 
+ \cdots  
. 
\end{eqnarray}%
It is clear that diagrams with arbitrary number of 
$\zeta $ vertices may be found in this series. 
Hence, despite the fact that the local source conditioned on the full 
radiated  
field is Gaussian, the radiated field itself is non-Gaussian; 
moreover, it has non-zero cumulants $C^{(n)}$ for all 
$n$. 
The number of connected diagrams is infinite, 
yet the diagram series retain certain ``tameness'': 
e.g., the number of diagrams contributing to each 
field cumulant, $C^{(n)}$, is still finite.

The final (and crucial) step leading to truly nontrivial  
series is combining nonlinearity and noise. 
Among the connected diagrams produced by relation (\ref{WickFieldReal}), 
we find all trees, both ``nonlinear'' (\ref{RealTrees}) and ``stochastic'' 
(\ref{TreesRealStoch}); on the top of that, we find  
a totally new class of 
netlike diagrams which are often called 
{\em diagrams with loops\/}:  
\begin{eqnarray} 
\ln \Phi (\zeta ) &=& \text{Trees} 
+  \frac{1}{2} \figPTbp
+ \frac{1}{2} \figPTbr
\nonumber  \\ && 
+ \figPTbs
\nonumber  \\ && 
+ \frac{1}{4} \figPTbt
+ \frac{1}{2} \figPTbu
+ \cdots . 
\end{eqnarray}%
Now, the number of connected diagrams contributing to any field 
cumulant is infinite.

\section{Nonlinear stochastic response problem}
The above CFT approach is a {\em radiation problem\/}, 
when one looks for properties of fields observable under given 
conditions; the quantum analog of it is, e.g.,  Glauber's quantum 
coherence  
theory \cite{Glauber}.
An alternative formulation is a {\em response problem\/}, 
when one is interested in the dependence of the field 
properties on external influences; in quantum mechanics, 
this is mainly associated with Kubo's linear response theory \cite{Kubo}.  
As was shown in \cite{OPO,Corresp}, the formal diagram solution 
to the  quantum nonlinear response problem has a structure 
identical to that of a classical nonlinear stochastic response problem. 
We therefore briefly outline how the above diagram expansions are 
adjusted to the ``response'' viewpoint. 

Following the spirit of Kubo's linear reaction theory \cite{Kubo},
we add a {\em given external source\/} to $s(r,t)$ in the 
Eqs (\ref{BasicEq}) and (\ref{BasicInt}), 
$s(r,t) \rightarrow s(r,t) + s\ext(r,t)$, and  
consider how the properties of the field $ \psi $ depend on an  
infinitesimally small perturbation $s\ext$. 
We define 
{\em nonlinear stochastic response functions\/} as,  
\begin{eqnarray} 
R^{(m,n)} \left ( 
x_1, \cdots, x_m 
;
x'_1, \cdots, x'_n 
\right ) = 
\frac{\partial ^n
\overline{ \psi (x_1) \cdots  \psi (x_m) }
}{
\partial s\ext(x'_1)
\cdots 
\partial s\ext(x'_n)
}
|_{s\ext = 0} 
.
\end{eqnarray}%
The response functions obey the causality 
condition,  
\begin{eqnarray} 
\showlabel{CausR}
R^{(m,n)} \left ( 
x_1, \cdots, x_m 
;
x'_1, \cdots, x'_n 
\right ) = 0, \ \ 
\max(t_1, \cdots, t_m) < 
\max(t'_1, \cdots, t'_n) . 
\end{eqnarray}%
For sufficiently small $s\ext$, 
(and assuming we are not at a phase transition 
point)  
\begin{eqnarray} 
\showlabel{TildeFByR}
\Phi \left ( 
\zeta ,s\ext
\right ) = \sum _{m,n=0}^{\infty} \frac{1}{m!n!} 
\zeta ^m R ^{(m,n)} s\ext ^n  , 
\end{eqnarray}%
where $\Phi \left ( 
\zeta ,s\ext
\right )$ is the characteristic functional 
of the field averages  in presence of the 
external source, so that $\Phi(\zeta ) = \Phi \left ( 
\zeta ,0
\right )$.

In diagram terms, adding the external source 
does not bring much new. 
One simply replaces $\chi ^{(1,0)} \rightarrow \chi ^{(1,0)} + s\ext$:  
\begin{eqnarray} 
 \showlabel{Resp}
\Phi \left ( 
\zeta ,s\ext
\right ) = \Phi ( 
\zeta ) _{\chi ^{(1,0)} \rightarrow \chi ^{(1,0)} + s\ext} . 
\end{eqnarray}%
This means considering two types 
of $\chi ^{(1,0)}$ vertices and summing 
all diagrams thus emerging. 
If for simplicity we assume that $\chi ^{(1,0)} = 0$,  
the connection between the ``radiation'' and ``response'' 
series becomes trivial: To find the latter, 
take the former with  
\begin{eqnarray} 
\dframe{4}{3}{\subp{0}{0}{%
\vrt{0}{0}
\dotted\linePT{0}{0}{4}{0}{\txt{-0.5}{1}{$x$}}
}
\tarr{0}{-2}{4}
}
= s\ext (x) . 
\end{eqnarray}%
Diagram expansions for the response 
functions are found by stripping the diagrams of  
the $\zeta $ and $\chi ^{(1,0)} = s\ext$ vertices, 
\begin{mathletters}
\showlabel{Rmn}
\begin{eqnarray} %
\nonumber  
R^{(1,1)}(x;x')&=& 
\figPTba + \figPTdy 
\\ && +  \frac{1}{2} \figPTdz + \cdots 
, 
\\
R^{(2,1)}(x,x';x'') &=& 
\figPTea + \figPTeb + \figPTec + \cdots 
, 
\\
R^{(1,2)}(x;x',x'') &=& 
\figPTed + \cdots
, 
\end{eqnarray}%
\end{mathletters}%
etc. 
It is easy to prove that  
the causality condition (\ref{CausR}) holds 
not only for an $R^{(m,n)}$ as a whole, 
but also for each particular diagram in its expansion.

\section{Constructing  an SDE 
for a given causal diagram series}
\label{merge}
Assume a diagram series is given 
(derived in a certain q-number approach, say). 
This series appears as 
a causal one, i.e., is generated by an expression 
like (\ref{WickFieldReal}), 
and the  causality conditions 
hold for all graphical elements. 
Formally, the generalised susceptibilities 
$\chi ^{(m,n)}$ then provide one with a complete and 
unambiguous description of the equivalent classical 
stochastic process. 
In practice, however, it would be more convenient to deal 
with an explicit SDE, written in terms of noise sources 
which are independent of the field. 
This leads us into what can be termed 
the inverse problem of the causal diagram techniques: 
How to explicitly write an SDE to which a causal diagram series 
corresponds?

With no stochasticity present, the relation between a 
 causal diagram series  
 and the corresponding DE 
is straightforward: 
A diagram series with only single-output 
vertices, $\chi ^{(1,n)}$, solves the equation, 
\begin{eqnarray} 
\showlabel{NonStoch}
{\cal L} \psi (x) = \sum_{n=1}^{\infty}  \frac{1}{n!} 
\int dx_1 \cdots dx_n
\chi ^{(1,n)}(x;x_1, \cdots , x_n) 
 \psi (x_1) \cdots  \psi (x_n) .  
\end{eqnarray}%
Basically, any SDE should look identical to this equation, 
with the only difference that (some of) the susceptibilities 
are random. 
It is clear that,  
\begin{eqnarray} 
\showlabel{PhiByPhi0}
\Phi (\zeta ) = \overline {\Phi _0 (\zeta )}, 
\end{eqnarray}%
where  $\Phi (\zeta )$ is the diagram series solving 
(\ref{NonStoch}) regarded as an SDE, 
$\Phi _0 (\zeta )$ is the series solving (\ref{NonStoch}) 
as a non-stochastic equation, and the upper bar here 
denotes averaging over the random susceptibilities 
in (\ref{NonStoch}). 
The key to the inverse problem is  in the relation between 
$\Phi _0 (\zeta )$ and $\Phi (\zeta )$. 

Consider, to begin with,  an example of a given 
random source. 
The SDE is then  
\begin{eqnarray} 
\showlabel{ExtFixed}
{\cal L} \psi (x) = s_0(x) . 
\end{eqnarray}%
For simplicity, we assume that $s_0$ is Gaussian with 
zero average, 
$\overline{s_0(x)} = 0$ 
 and  
$\overline{s_0(x) s_0(x')} = \chi ^{(2,0)}(x,x')$. 
Then, 
\begin{eqnarray} 
\Phi _0 (\zeta ) &=& \exp \figPTai , 
\\
\Phi (\zeta ) &=& \exp\left ( 
\frac{1}{2} \figPTbh 
\right ). 
\end{eqnarray}%
On the other hand,  
expanding  $\Phi _0 (\zeta )$ in a diagram series 
and performing the averaging as required by 
(\ref{PhiByPhi0}) in each of the diagrams 
separately, we find:  
\begin{eqnarray} 
\Phi (\zeta )  &=& 1 + 
\overline{\figPTai } + 
 \frac{1}{2} \overline{\figPTdo } + 
 \frac{1}{6} \overline{\figPTdp } + 
 \frac{1}{24} \overline{\figPTdq } + \cdots 
\nonumber  \\ 
&=& 
1 + 
0 + 
 \frac{1}{2} \figPTbh  + 
0 + 
 \frac{1}{8} \figPTdr  + \cdots .
\end{eqnarray}%
We see that  the graphical operation reflecting $s_0$ 
becoming random and Gaussian 
is a pairwise {\em merging\/} of the $\chi ^{(1,0)}$ vertices 
into $\chi ^{(2,0)}$ vertices; 
diagrams with an odd number  of the $\chi ^{(1,0)}$ vertices 
become zero. 
Note that whereas itself the merging of the vertices 
reflects stochasticity, 
the fact that 
it is pairwise is clearly due to the Gaussian statistics: 
if $s_0$ were non-Gaussian, then  
non-zero $\chi ^{(m,0)}$ vertices would each result from 
merging of $m$ $\chi ^{(1,0)}$ vertices. 

Consider now an SDE with 
a multiplicative noise,  
\begin{eqnarray} 
\showlabel{Mult}
{\cal L} \psi (x) = \int dx'  \eta (x;x')  \psi (x') + s'(x) , 
\end{eqnarray}%
where $\eta (x;x')$ is random and  
$s'(x)$ contains non-stochastic terms  
(it may also contain other noise sources provided  
they are not correlated with $\eta $). 
For simplicity, we again assume that 
$\eta (x;x')$ is Gaussian and  
$\overline{\eta (x;x')} = 0$, so that it is 
specified by the average  
$\overline{\eta (x_1;x'_1)\eta (x_2;x'_2)}$. 
The diagram series for equation (\ref{Mult}) can be found 
by averaging those for an equation 
with a linear susceptibility 
$\chi ^{(1,1)}(x;x') =  \eta (x;x')$. 
This is most intuitive if done directly in the generating expression 
(\ref{WickFieldReal}): 
\begin{eqnarray} 
\Phi (\zeta ) &=& 
\exp\left ( 
\frac{\partial }{\partial  \psi }
G
\frac{\partial }{\partial \alpha }
\right )
\overline{\exp \figPTak }
\exp  \left ( 
\zeta  \psi 
\right )  
S'\left ( 
\alpha |  \psi 
\right ) 
|_{\alpha =  0, \psi =  \psi _ 0} 
\nonumber  \\ &=& 
\exp\left ( 
\frac{\partial }{\partial  \psi }
G
\frac{\partial }{\partial \alpha }
\right )
\exp \left ( 
 \frac{1}{4} \figPTds
\right )
\exp  \left ( 
\zeta  \psi 
\right )  
S'\left ( 
\alpha |  \psi 
\right ) 
|_{\alpha =  0, \psi =  \psi _ 0} 
, 
\end{eqnarray}%
where  $S'\left ( 
\alpha |  \psi 
\right )$ contains vertices originating in $s'(x)$,  and 
the following graphical notation is used,  
\begin{eqnarray} 
\figPTo  &=& \psi (x) , 
\showlabel{PsiLine}
\\
\figPTp &=& \alpha (x) , 
\showlabel{AlphaLine}
\\
\figPTak &=& \alpha \chi ^{(1,1)} \psi , \\
 \figPTds &=& \alpha ^2 \chi ^{(2,2)} \psi ^2. 
\\
\showlabel{MultCorr}
\figPTdv &=& \chi ^{(2,2)}(x_1,x_2;x'_1,x'_2)
= 
\overline{\eta (x_1;x'_1)\eta (x_2;x'_2)}
+ \overline{\eta (x_1;x'_2)\eta (x_2;x'_1)} .
\end{eqnarray}%
(The quantities $\alpha ^m \chi ^{(m,n)}  \psi ^n$ 
are defined in Eq.\ (\ref{ChiGen}). 
Note that we have departed from the case 
of only the
five  susceptibilities shown explicitly in Eqs. 
(\ref{FullSource}) being non-zero.)
Hence the graphical representation 
of the linear susceptibility 
becoming random is pairs of the 
$\chi ^{(1,1)}$ vertices merging 
into 
 the $\chi ^{(2,2)}$ vertices, e.g.,  
\begin{eqnarray} 
 \frac{1}{2} \figPTdt 
\rightarrow
 \frac{1}{4} \figPTdu .
\end{eqnarray}%
Similarly, if 
a certain pair of susceptibilities 
($\chi ^{(1,0)}$ and $\chi ^{(1,2)}$, say)
become random and correlated (implying Gaussian), then 
averaging $\Phi_0$ we get, 
\begin{eqnarray} 
\Phi (\zeta ) &=& \exp\left ( 
\frac{\partial }{\partial  \psi }
G
\frac{\partial }{\partial \alpha }
\right )
\overline{\exp \left ( 
\figPTr +  \frac{1}{2} \figPTdw 
\right )}
\exp  \left ( 
\zeta  \psi 
\right )  
S'\left ( 
\alpha |  \psi 
\right ) 
|_{\alpha =  0, \psi =  \psi _ 0} 
\nonumber  \\ &=& 
\exp\left ( 
\frac{\partial }{\partial  \psi }
G
\frac{\partial }{\partial \alpha }
\right )
\exp \left ( 
 \frac{1}{2} \figPTbg + 
 \frac{1}{48} \figPTdx + 
 \frac{1}{4} \figPTds 
\right ) 
\\ \nonumber && \times 
\exp  \left ( 
\zeta  \psi 
\right )  
S'\left ( 
\alpha |  \psi 
\right ) 
|_{\alpha =  0, \psi =  \psi _ 0} 
, 
\end{eqnarray}%
where  
\begin{eqnarray} 
\figPTr  &=& \alpha \chi ^{(1,0)}, \\
\figPTdw &=& \alpha \chi ^{(1,2)} \psi ^2, \\
\figPTbg &=& \alpha ^2 \chi ^{(2,0)}, \\
\figPTdx &=& \alpha ^2 \chi ^{(2,4)} \psi ^4, \\
 \figPTds &=& \alpha ^2 \chi ^{(2,2)} \psi ^2, 
\end{eqnarray}%
and 
\begin{mathletters}
\showlabel{CrossCorr}
\begin{eqnarray} 
\chi ^{(2,0)}(x_1,x_2) &=& 
\overline{\chi ^{(1,0)}(x_1)\chi ^{(1,0)}(x_2)}, 
\\  
\chi ^{(2,4)}(x_1,x_2;x'_1,x'_2,x'_3,x'_4) &=& 
\text{Symm}\, \overline{\chi ^{(1,2)}(x_1;x'_1,x'_2)
\chi ^{(1,2)}(x_2;x'_3,x'_4)}, 
\\     
\chi ^{(2,2)}(x_1,x_2;x'_1,x'_2) &=& 
\text{Symm}\, \overline{\chi ^{(1,1)}(x_1;x'_1)
\chi ^{(1,1)}(x_2;x'_2)}    .
\end{eqnarray}%
\end{mathletters}%
Symm here denotes summation over all different terms 
obtained by permutations of the input and output arguments, 
cf (\ref{MultCorr}). 
We see that whereas randomness results  in 
self-mergings 
($\chi ^{(1,0)} + \chi ^{(1,0)} \rightarrow \chi ^{(2,0)}$
and 
$\chi ^{(1,2)} + \chi ^{(1,2)} \rightarrow \chi ^{(2,4)}$), 
correlations manifest themselves as cross-mergings  
($\chi ^{(1,0)} + \chi ^{(1,2)} \rightarrow \chi ^{(2,2)}$).

This way, if we take equation (\ref{NonStoch}), derive 
a diagram series for it, and then assume that the 
susceptibilities $\chi ^{(1,n)}$ are random, the new series 
is found from the initial series by merging certain vertices 
in the diagrams. 
Hence in order to recover the initial non-stochastic equation, 
 the merged vertices should be factorised into the products 
of the initial vertices. 
Formally, this means solving functional equations like 
(\ref{MultCorr}) or (\ref{CrossCorr}), 
with given stochastic vertices, 
in order to find statistics of the susceptibilities in (\ref{NonStoch}). 
This is nothing but the well-known problem of moments 
in probability theory. 

It is obvious that there exist causal diagram series that 
do not correspond to any SDE in the true meaning of the 
word.  
E.g., $\chi ^{(2,0)}(x;x') = -  \delta (x-x')$ 
would require $\overline{s_0(x) s_0(x')} = -  \delta (x-x')$; 
this is certainly impossible in probability theory. 
We therefore have to adopt the 
concept of 
pseudo-probability \cite{QNoise} and consider 
pseudo-stochastic 
differential equations  (PSDE) as well as stochastic ones.  
(This is anyway inevitable in quantum stochastics 
since the measure in Feynman path integrals is as a rule 
nonpositive.) 
Even with this generalisation, it is not clear if the 
inverse problem has a general solution consistent 
with the causality conditions for the susceptibilities. 

However, this clearly is the case for an important class 
of problems, namely,  
local Markovian problems.  
In terms of the susceptibilities, this means 
that they are non-zero only if all their arguments 
(both input and output) coincide. 
E.g., let $\chi ^{(2,2)}(x_1,x_2;x_1',x_2') 
= \chi   
\delta (x_1-x_1')\delta (x_2-x_2')\delta (x_1-x_2)$, where  
$\chi $ is a real constant, positive or negative. 
Then, the inverse problem is solved by 
$\chi ^{(1,1)}(x;x') = \sqrt{|\chi |}  \,
\eta (x) \delta (x-x')$, where 
for $\chi >0$  $\eta  (x)$ is a standardised Gaussian 
$ \delta $-correlated noise source, 
$\overline{\eta (x)\eta  (x')} =  \delta (x-x')$; 
whereas for $\chi < 0$  $\eta (x)$ is a standardised Gaussian 
$ \delta $-correlated {\em pseudo-stochastic\/} source, 
$\overline{\eta (x)\eta  (x')} =  - \delta (x-x')$. 
Note that one can equally find an alternative solution 
using factorisation (\ref{CrossCorr}). 
This shows that a solution to an inverse problem 
is in general non-unique.  

\appendix
\section*{} 
We now show that relation (\ref{WickFieldReal}) can 
formally be expanded in a series of the causal diagrams. 
To this end, we re-write it as,  
\begin{eqnarray} 
\Phi (\zeta ) &=&  \sum_{p,q,r,s,t ,u = 1}^{\infty}
\frac{1}{p!q!r!s!t !u!} 
\exp {\cal D} \figPTq ^p \figPTr ^q  \figPTak ^r
\times \nonumber \\ && 
\left ( 
 \frac{1}{2} 
\figPTbg 
\right )^s
\left ( 
 \frac{1}{2} 
\figPTbk 
\right )^t
\left ( 
 \frac{1}{2} 
\figPTdw
\right )^u
| _{\alpha  =  \psi  = 0}, 
\showlabel{PhiFull}
\end{eqnarray}%
where  
\begin{eqnarray} 
{\cal D} = {\cal D}_1 + {\cal D}_2 =
\frac{\partial  }{\partial  \psi } \psi _0 
+ 
\frac{\partial }{\partial  \psi }
G
\frac{\partial }{\partial  \alpha }, 
\end{eqnarray}%
and  
\begin{eqnarray} 
\figPTq &=& \zeta  \psi , \\
\figPTbk &=& \alpha ^2 \chi ^{(2,1)} \psi . \\
\end{eqnarray}%
The rest of the graphical notation used in 
(\ref{PhiFull}) was introduced in Chapter \ref{merge}.
Note that we have returned to the case 
of only the
five  susceptibilities shown explicitly in Eqs. 
(\ref{FullSource}) being non-zero.

Rewriting the product of connected diagrams in 
(\ref{PhiFull}) as a  disconnected diagram, 
the general term in the series (\ref{PhiFull}) is found 
to be, 
\begin{eqnarray} 
\showlabel{GenTerm}
\text{const} \times 
\exp {\cal D} 
\left \{ 
D_{pqrstu}
\right \}
|_{\alpha  =  \psi  = 0}. 
\end{eqnarray}%
Here $\left \{ 
D_{pqrstu}
\right \}
$ is a disconnected diagram 
with $p+q+r+s+t +u$ connected components, 
where $\figPTq $ is repeated $p$ times, 
$\figPTr  $ is repeated $q$ times, etc. 
To find a graphical representation of the 
operation $\exp {\cal D} 
\left \{ 
D_{pqrstu}
\right \}
|_{\alpha  =  \psi  = 0}
$, we notice that,  
\begin{eqnarray} 
\left \{ 
D_{pqrstu}
\right \}
 &=& 
\int 
dx_1\cdots dx_m 
dx'_1\cdots dx'_n
\nonumber  \times \\ &&
 \psi (x_1)\cdots  \psi (x_m )
\alpha (x'_1)\cdots \alpha (x'_n)
\Xi (
x'_1,\cdots ,x'_n ;
x_1,\cdots , x_m 
) ,
\showlabel{Dpqrstu}
\end{eqnarray}%
where $m = p+r+t+2u$, 
$n = q + r + 2s + 2t + u$, 
and 
$\Xi (
x'_1,\cdots ,x'_n ;
x_1,\cdots , x_m 
) 
$ is a product of the generalised susceptibilies and $\zeta $s; 
$x_1,\cdots , x_m$ and 
$x'_1,\cdots , x'_n$ are, respectively, 
input and output arguments, each belonging to  
a certain susceptibility factor (or to a $\zeta $)
in this product. 
$\Xi $ contains neither $\alpha $s nor $ \psi $s, 
hence we have to understand only how $\exp {\cal D} $ 
acts on the product of the $\alpha $s and $ \psi $s. 

Consider firstly the operator $\exp {\cal D} _1$, 
where ${\cal D} _1 =  \frac{\partial }{\partial  \psi }\psi _0 $. 
It is readily seen that  
\begin{eqnarray} 
\showlabel{D1Base}
{\cal D} _1   \psi (x_1)\cdots  \psi (x_m )
= 
\sum _{l = 1}^{m} 
 \psi (x_1)\cdots  \psi (x_{l-1} )
\psi _0 (x_l)
 \psi (x_{l+1})\cdots  \psi (x_m ). 
\end{eqnarray}%
That is, ${\cal D} _1 $ turns a product of $m$ $ \psi $s 
into $m$ terms, in each of which one factor 
$ \psi $ is replaced by $ \psi _0$. 
Then, $ \frac{1}{2!} {\cal D} _1 ^2$ will turn the 
same product into $m(m-1)/2$ terms, 
in each of which two factors 
$ \psi $ are replaced by $ \psi _0$s; 
the factor $ \frac{1}{2!}$ exactly compensates 
for two possible orders of replacement leading 
to each term. 
It is now easy to realise (and prove by induction) that 
$ \frac{1}{k!} {\cal D} _1 ^k$ will turn the 
product of $m$ $ \psi $s 
into $ \frac{m!}{k!  (m - k)!}$ different terms, 
in each of which k factors 
$ \psi $ are replaced by $ \psi _0$. 
At $k = m$, we shall find one term with 
all $ \psi $s replaced 
by $ \psi _0$s,  
$ \frac{1}{m!} {\cal D} _1 ^m 
\psi (x_1)\cdots  \psi (x_m ) = 
\psi _0 (x_1)\cdots  \psi _0 (x_m )
$, and  ${\cal D} _1 ^k 
\psi (x_1)\cdots  \psi (x_m ) = 0, k > m$. 
Thus,  
\begin{eqnarray} 
&& \exp {\cal D}_1 \psi (x_1)\cdots  \psi (x_m ) = 
\left ( 
1 +  {\cal D} _1 + \frac{1}{2!} {\cal D} _1 ^2 
+ \cdots + \frac{1}{m!} {\cal D} _1 ^m
\right ) \psi (x_1)\cdots  \psi (x_m )  
\nonumber  \\ &&
= \text{Sum of all different products of $k $ $ \psi _0$s 
and $m - k$ $ \psi $s, $0 \leq k \leq m $}. 
\end{eqnarray}%
Note that this result is based primarily on relation 
(\ref{D1Base}); 
the rest of the derivation is mere combinatorics. 

Similarly, how the operator $\exp {\cal D} _2$, 
where ${\cal D} _2 =  
\frac{\partial }{\partial  \psi }
G
\frac{\partial }{\partial  \alpha }$, acts on a product of 
$\alpha $s and $ \psi $s, is based primarily on the relation,  
\begin{eqnarray} 
{\cal D} _2 
\psi (x_1)\cdots  \psi (x_m )
\alpha (x'_1)\cdots \alpha (x'_n) &=& 
\sum _{l=1}^{m}
\sum _{j=1}^{n}
 G(x_l;x_j)
\psi (x_1)\cdots  \psi (x_{l-1} )
 \psi (x_{l+1})\cdots  \psi (x_m )
\nonumber  \times \\ &&
\alpha (x'_1)\cdots  \alpha (x'_{j-1} )
 \alpha (x'_{j+1})\cdots  \alpha (x'_n )
. 
\end{eqnarray}%
This relation reads: in the product 
$\psi (x_1)\cdots  \psi (x_m )
\alpha (x'_1)\cdots \alpha (x'_n)  
$, select a pair $ \psi (x_l), \alpha (x'_j)$, 
and replace it by $G(x_l;x'_j)$; 
${\cal D} _2 
\psi (x_1)\cdots  \psi (x_m )
\alpha (x'_1)\cdots \alpha (x'_n)$ is then found as a sum 
of all different terms obtained this way. 
Then, $ \frac{1}{k!} {\cal D} _2 ^k
\psi (x_1)\cdots  \psi (x_m )
\alpha (x'_1)\cdots \alpha (x'_n)$ 
is equal to the sum of all different terms, found by 
replacing  $k$ 
such pairs by the $G$s in the initial product; 
if $k > \min (m,n)$, one finds zero. 
Finally,  
\begin{eqnarray} 
&& \exp{\cal D} _2 \psi (x_1)\cdots  \psi (x_m )
\alpha (x'_1)\cdots \alpha (x'_n) 
\nonumber  \\ && 
= \text{Sum of all different products of $k$ $G$s, 
$m - k $ $ \psi $s 
and $n - k$ $ \alpha $s, $0 \leq k \leq \min (m,n) $}. 
\end{eqnarray}%
(This clearly follows the pattern of Wick's theorem for 
bosonic operators, replacing operator pairs by 
the ``junction'' $G$, cf \cite{Akhiezer,Hori}.)
Now there is no problem in understanding
 the action of the operator $\exp {\cal D} = 
\exp ({\cal D}_1 + {\cal D}_2)$:
\begin{eqnarray} 
&& \exp{\cal D} \psi (x_1)\cdots  \psi (x_m )
\alpha (x'_1)\cdots \alpha (x'_n) 
\nonumber  \\ && 
= \text{Sum of all different products of $k$ $G$s, 
$l$ $ \psi _0$s,
$m - k - l$ $ \psi $s 
and $n - k$ $ \alpha $s}, 
\nonumber  \\ && 
0 \leq k \leq n,  0 \leq l+k \leq m  . 
\end{eqnarray}%

The graphical recipe corresponding to this relation is 
quite transparent: (i) represent the product 
$\psi (x_1)\cdots  \psi (x_m )
\alpha (x'_1)\cdots \alpha (x'_n)$ as a disconnected 
diagram with 
$m$ elements (\ref{PsiLine}) 
and $n$ elements (\ref{AlphaLine}), then 
(ii) write a sum of all different diagrams obtainable  
by replacing randomly chosen $ \psi $s by $ \psi _0$s   
and pairs $\alpha , \psi $ by $G$s, 
\begin{eqnarray} 
\figPTo &\rightarrow &
\vframe{4.5}{3.5}{0}{0}{%
\lnd{0}{0}{4}{}{$\ \ \ x$}{}
\tarr{0.25}{-2}{4}
}  , \\
\figPTs &\rightarrow & \figPTt . 
\end{eqnarray}%
For example, 
\begin{eqnarray} 
\exp {\cal D}  \ 
\alpha (x)  \psi (x')
&=& 
\exp {\cal D}  \ 
\figPTs = \figPTs + \figPTt , \\
\exp {\cal D} \  
\alpha (x)  \psi (x')   \psi (x'')
&=& 
\exp {\cal D} \  
\figPTu 
\nonumber  \\ &=& 
 \figPTu + \figPTv  + \figPTw + 
\nonumber  \\ && 
\figPTbv  + \figPTbw , 
\end{eqnarray}%
Then, setting $\alpha  =  \psi  = 0$ cancels   
all terms with dashed lines, 
e.g.,  
\begin{mathletters}
\showlabel{WickRealGraph}
\begin{eqnarray} 
\exp {\cal D}  \ 
\figPTs |_{\alpha  =  \psi  = 0} 
&=& 
\figPTt , \\
\exp {\cal D} \  
\figPTu  |_{\alpha  =  \psi  = 0}
&=&
\figPTbv  + \figPTbw , \\  
\exp {\cal D} \  
\figPTbx  |_{\alpha  =  \psi  = 0}
&=&
\figPTby  + \figPTbz .  
\end{eqnarray}%
\end{mathletters}%
Thus, the graphical recipe for calculating 
$\exp{\cal D}\,  \psi (x_1)\cdots  \psi (x_m )
\alpha (x'_1)\cdots \alpha (x'_n) |_{\alpha  =  \psi  = 0}
$ is that for calculating 
$\exp{\cal D} \psi (x_1)\cdots  \psi (x_m )
\alpha (x'_1)\cdots \alpha (x'_n) 
$, plus (iii) retain only terms without dashed lines. 
Note that the number of resulting 
$ \psi _0$s and $G$s in all such terms 
is fixed ($n$ and $m-n$, respectively), 
while their arguments vary from term 
to term. 

It is now obvious that 
 to calculate (\ref{GenTerm}) graphically, 
one should remove pairs of dashed lines 
$\alpha ,  \psi $ in the diagram 
$\left \{ 
D_{pqrstu}
\right \}
$, installing instead solid lines connecting 
inputs of the vertices to outputs, so that no free 
outputs remain; 
then, all remaining dashed  $ \psi $s should be replaced 
by solid $ \psi _0$s, connected to inputs of the vertices.   
Terms where the number of outputs in $\left \{ 
D_{pqrstu}
\right \}
$ exceeds the number of inputs do not contribute.
Each terms in series (\ref{PhiFull}) is thus equal to 
either a sum of causal diagrams or zero, 
e.g., 
\begin{eqnarray} 
\exp {\cal D} \figPTr |_{\alpha = \psi = 0} &=&  0 , 
\\
\exp {\cal D} \figPTq |_{\alpha = \psi = 0} &=& \figPTac  ,
 \\
 \frac{1}{2} \exp {\cal D} \figPTq^2 |_{\alpha = \psi = 0} 
&=& 
 \frac{1}{2} \exp {\cal D} \figPTad |_{\alpha = \psi = 0} 
= 
\frac{1}{2} \figPTaf  ,
 \\
\exp {\cal D} \figPTq \figPTr |_{\alpha = \psi = 0} 
&=& 
\exp {\cal D} \figPTag |_{\alpha = \psi = 0} 
= 
\figPTai  , 
 \\
 \frac{1}{4} \exp {\cal D} \figPTq^2 \figPTr^2 
|_{\alpha = \psi = 0} 
&=& 
 \frac{1}{4} \exp {\cal D} \figPTca |_{\alpha = \psi = 0} 
\nonumber  \\ &=& 
 \frac{1}{4} \figPTcb +  \frac{1}{4} \figPTcc 
\nonumber  \\ &=& 
 \frac{1}{2} \figPTcb . 
\end{eqnarray}%
The last example shows that, in general, 
  not all diagrams obtained from 
a particular $\left \{ 
D_{pqrstu}
\right \}
$ are different  (unlike in sums (\ref{WickRealGraph})); 
this is due to possible symmetry of the vertex product
$\Xi$.
All such equal diagrams  count when calculating 
the coefficient with which a diagram appears in the series.  
It is obvious that any given causal diagram, 
connected or disconnected,  
 is produced by 
a unique  
term in series (\ref{PhiFull}), namely, 
by the one containing the same set of vertices as the 
diagram sought.  
It is then follows  that all complete causal diagrams
are found in the resulting diagram series.

\end{document}